\documentclass[12pt,tightenlines,eqsecnum,floats,aps,amsmath,amssymb,nofootinbib,prd,showpacs]{revtex4}

\usepackage{setspace}
\usepackage{amsmath,amssymb,amsfonts,amsthm}
\usepackage{graphicx}
\usepackage{enumerate} 
\usepackage{colordvi} 

\def\be{\begin{equation}}
\def\ee{\end{equation}}
\def\ba{\begin{eqnarray}}
\def\ea{\end{eqnarray}}

\def\p{\partial}
\def\V{\mathcal{C}}
\def\H{{\cal H}}

\def\Hp{\H_{\rm phy}}
\def\F{\mathcal{F}}
\def\h{\hat }
\def\b{\bar}

\def\cm{\rm cm}

\def\pphi{p_{(\phi)}}

\def\lp{{\ell}_{\rm Pl}}
\def\lo{{\ell}_o}
\def\R{\mathbb{R}}
\def\S{\mathbb{S}}

\def\B{\mathbb{B}}
\def\E{E}

\def\q{{}^o\!q}
\def\e{{}^o\!e}

\def\hor{{\rm hor}}
\def\rcr{\rho_{\mathrm{crit}}}
\def\rmin{\rho_{\mathrm{min}}}
\def\rmax{\rho_{\mathrm{max}}}
\def\b{$\bullet\,\,\,\, $}
\def\f{\frac}
\def\dd{\textrm{d}}

\def\ul{\underline}
\def\WDW{WDW\,\,}

\def\szm{\mathcal{I}^{o-}}

\def\szpr{\mathcal{I}^{o+}_{\rm R}}

\def\szmr{\mathcal{I}^{o-}_{\rm R}}
\def\szml{\mathcal{I}^{o-}_{\rm L}}
\def\sbpr{{\mathcal{I}}^{+}_{\rm R}}
\def\spr{\mathcal{I}^{+}_{\rm R}}
\def\spl{\mathcal{I}^{+}_{\rm L}}

\def\sm{\mathcal{I}^{-}}
\def\sp{\mathcal{I}^{+}}

\newcommand{\ket}[1]{\ensuremath{|#1\rangle}}
\def\k{\kappa}

\newcommand{\fs}[2]{{\textstyle\frac{#1}{#2}}} 

\usepackage{enumerate}

\usepackage{colordvi}
\newcounter{mnotecount}[section]

\newcommand{\comment}[1]{}

\begin{document}

\title{Quantum Space-times:\\
Beyond the Continuum of Minkowski and Einstein}

\author{Abhay Ashtekar}
\email{ashtekar@gravity.psu.edu} \affiliation{Institute for
Gravitation and the Cosmos, Physics Department, Penn State,
University Park, PA 16802, U.S.A.}

\begin{abstract}

In general relativity space-time ends at singularities. The big bang
is considered as the Beginning and the big crunch, the End. However
these conclusions are arrived at by using general relativity in
regimes which lie well beyond its physical domain of validity.
Examples where detailed analysis is possible show that these
singularities are naturally resolved by quantum geometry effects.
Quantum space-times can be vastly larger than what Einstein had us
believe. These non-trivial space-time extensions enable us to answer
of some long standing questions and resolve of some puzzles in
fundamental physics. Thus, a century after Minkowski's revolutionary
ideas on the nature of space and time, yet another paradigm shift
appears to await us in the wings.

\end{abstract}

\pacs{04.60.Kz,04.60Pp,98.80Qc,03.65.Sq}

\maketitle

\section{Introduction}
\label{s1}

A hundred years ago Hermann Minkowski fused space and time into a
smooth 4-dimensional continuum. Remarkably, this continuum
---the Minkowski space-time--- still serves as the arena for all
non-gravitational interactions both in classical \emph{and} quantum
physics. Time is no more absolute. Whereas in Newtonian physics
there is a unique 3-plane through each space-time point representing
space, now there is a unique cone, spanned by light rays passing
through that point. The constant time plane curls up into a 2
sheeted cone that separates the region which is causally connected
with the point from the region which is not. This causality dictates
the propagation of physical fields in classical physics, and the
commutation relations and uncertainty relations between field
operators in quantum physics. With the demise of absolute
simultaneity, Newtonian ideas are shattered. The world view of
physics is dramatically altered.

However, as in Newtonian physics, there is still a fixed space-time
which serves as the arena for all of physics. It is the stage on
which the drama of evolution unfolds. Actors are particles and
fields. The stage constrains what the actors can do. The Minkowski
metric dictates the field equations and restricts the forms of
interaction terms in the action. But the actors cannot influence the
stage; Minkowskian geometry is immune from change. To incorporate
the gravitational force, however, we had to abandon this cherished
paradigm. We follow Einstein and encode gravity in the very geometry
of space-time. Matter curves space-time. The space-time metric is no
longer fixed. There is again a dramatic paradigm shift. However, we
continue to retain one basic feature of Newtonian and Minkowskian
frameworks: space-time is still represented by a smooth continuum.

This is not uncommon: New paradigms are often created by
abandoning one key feature of the older paradigm but retaining
another. But global coherence of the description of Nature is a
huge burden and such a strategy often leads to new tensions. For
example, to achieve compatibility between mechanics and Maxwellian
electrodynamics, Einstein abandoned absolute simultaneity but
retained the idea that space and time are fixed, unaffected by
matter. The strategy worked brilliantly. Not only was the new
mechanics compatible with Maxwell's theory but it led to deep,
unforeseen insights. Energy and mass are simply two facets of the
same physical attribute, related by $E=mc^2$; electric and
magnetic fields $\vec{E}, \vec{B}$ are but two projections of an
electromagnetic field tensor $F_{ab}$; in a quantum theory of
charged particles, each particle must be accompanied by an
anti-particle with opposite charge. However, the new mechanics
flatly contradicted basic tenets of Newton's theory of
gravitation. To restore coherence of physics, one has to abandon
the idea that space-time is fixed, immune to change. One had to
encode gravity into the very geometry of space-time, thereby
making this geometry dynamical.

Now the situation is similar with general relativity itself.
Einstein abandoned the tenet that geometry is inert and made it a
physical entity that interacts with matter. This deep paradigm
shift again leads to unforeseen consequences that are even more
profound. Thanks to this encoding, general relativity predicts
that the universe began with a big bang; that heavy stars end
their lives through a gravitational collapse to a black hole;
that ripples in the space-time curvature propagate as
gravitational waves carrying energy-momentum. However, general
relativity continues to retain the Newtonian and Minkowskian
premise that space-time is a smooth continuum. As a consequence,
new tensions arise.

In Newtonian or Minkowskian physics, a given physical field could
become singular at a space-time point. This generally implied that
the field could not be unambiguously evolved to the future of that
point. However, this singularity had no effect on the global arena.
Since the space-time geometry is unaffected by matter, it remains
intact. Other fields could be evolved indefinitely. Trouble was
limited to the one field which became ill behaved. However, because
gravity is geometry in general relativity, when the gravitational
field becomes singular, the continuum tares and the space-time
itself ends.  There is no more an arena for other fields to live in.
All of physics, as we know it, comes to an abrupt halt. Physical
observables associated with both matter and geometry simply diverge
signalling a fundamental flaw in our description of Nature. This is
the new quandary.

When faced with deep quandaries, one has to carefully analyze the
reasoning that led to the impasse. Typically  the reasoning is
flawed, possibly for subtle reasons. In the present case the
culprit is the premise that general relativity ---with its
representation of space-time as a smooth continuum--- provides an
accurate description of Nature arbitrarily close to the
singularity. For, general relativity completely ignores quantum
effects and, over the last century, we have learned that these
effects become important in the physics of the small. They should
in fact be \emph{dominant} in parts of the universe where matter
densities become enormous. Thus there is no reason to trust the
predictions of general relativity near space-time singularities.
Classical physics of general relativity does come to a  halt at
the big-bang and the big crunch. But this is not an indication of
what \emph{really} happens because use of general relativity near
singularities is an extrapolation which has no physical
justification whatsoever. We need a theory that incorporates not
only the dynamical nature of geometry but also the ramifications
of quantum physics. We need a quantum theory of gravity, a new
paradigm.


These considerations suggest that singularities of general
relativity are perhaps the most promising gates to physics beyond
Einstein. They provide a fertile conceptual and technical ground in
our search of the new paradigm. Consider some of the deepest
conceptual questions we face today: the issue of the Beginning and
the end End; the arrow of time; and the puzzle of black hole
information loss. Their resolutions hinge on the true nature of
singularities. In my view, considerable amount of contemporary
confusion about such questions arises from our explicit or implicit
insistence that singularities of general relativity are true
boundaries of space-time; that we can trust causal structure all the
way to these singularities; that notions such as event horizons are
absolute even though changes in the metric in a Planck scale
neighborhood of the singularity can move event horizons dramatically
or even make them disappear altogether \cite{ph}.

Over the last 2-3 years several classically singular space-times
have been investigated in detail through the lens of loop quantum
gravity (LQG) \cite{alrev,crbook,ttbook}. This is a non-perturbative
approach to the unification of general relativity and quantum
physics in which one takes Einstein's encoding of gravity into
geometry seriously and elevates it to the quantum level. One is thus
led to build quantum gravity using \emph{quantum} Riemannian
geometry \cite{almmt,rs,al5,alvol}. Both geometry and matter are
\emph{dynamical} and described \emph{quantum mechanically} from the
start. In particular, then, there is no background space-time. The
kinematical structure of the theory has been firmly established for
some years now. There are also several interesting and concrete
proposals for dynamics (see, in particular
\cite{alrev,crbook,ttbook,spinfoam-rev}). However, in my view there
is still considerable ambiguity and none of the proposals is fully
satisfactory. Nonetheless, over the last 2-3 years, considerable
progress could be made by restricting oneself to subcases where
detailed and explicit analysis is possible
\cite{mb1,mb-rev,aa-rev,ab1,kv,atv}. These `mini' and `midi'
superspaces are well adapted to analyze the deep conceptual tensions
discussed above. For, they consider the most interesting of
classically singular space-times ---Friedman-Robertson-Walker (FRW)
universes with the big bang singularity and black holes with the
Schwarzschild-type singularity--- and analyze them in detail using
symmetry reduced versions of loop quantum gravity. In all cases
studied so far, classical singularities are naturally resolved and
\emph{the quantum space-time is vastly larger than what general
relativity had us believe.} As a result, there is a new paradigm to
analyze the old questions.

The purpose of this article is to summarize these developments,
emphasizing the conceptual aspects%
\footnote{Thus I will not include any derivations but instead
provide references where the details can be found.}
from an angle that, I hope, will interest not only physicists but
especially philosophers and historians of science. We will see that
some of the long standing questions can be directly answered, some
lose their force in the new paradigm while others have to be
rephrased.

This chapter is organized as follows. In section \ref{s2} I will
discuss cosmological singularities and in \ref{s3} the black hole
singularities. In each case I will discuss examples of fundamental
open issues and explain their status in the corresponding models. We
will see that quantum geometry has unexpected ramifications that
either resolve or significantly alter the status of these issues.
Finally in section \ref{s4} I will summarize the outlook and discuss
some of the fresh challenges that the new paradigm creates.

\section{Quantum Nature of the Big Bang}
\label{s2}

\subsection{Issue of the Beginning and the End}
\label{s2.1}

Over the history of mankind, cosmological paradigms have evolved
in interesting ways. It is illuminating to begin with a long range
historical perspective by recalling paradigms that seemed obvious
and most natural for centuries only to be superseded by radical
shifts.

Treatise on Time, the Beginning and the End date back at least
twenty five centuries. Does the flow of time have an objective,
universal meaning beyond human perception? Or, is it fundamentally
only a convenient, and perhaps merely psychological, notion? Did the
physical universe have a finite beginning or has it been evolving
eternally? Leading thinkers across cultures meditated on these
issues and arrived at definite but strikingly different answers. For
example, in the sixth century BCE, Gautama Buddha taught that `a
period of time' is a purely conventional notion, time and space
exist only in relation to our experience, and the universe is
eternal. In the Christian thought, by contrast, the universe had a
finite beginning and there was debate whether time represents
`movement' of  bodies or if it flows only in the soul. In the fourth
century CE, St. Augustine held that time itself started with the
world.

Founding fathers of modern Science from Galileo to Newton continued
to accept that God created the universe. Nonetheless, their work led
to a radical change of paradigm. Before Newton, boundaries between
the absolute and the relative, the true and the apparent and the
mathematical and the common were blurry. Newton rescued time from
the psychological \emph{and} the material world and made it
objective and absolute. It now ran uniformly from the infinite past
to the infinite future. This paradigm became a dogma over centuries.
Philosophers often used it to argue that the universe itself
\emph{had} to be eternal. For, as Immanuel Kant emphasized,
otherwise one could ask ``what was there before?''

General relativity toppled this Newtonian paradigm in one fell
swoop. Now the gravitational field is encoded in space-time
geometry. Since geometry is a dynamical, physical entity, it is now
perfectly feasible for the universe to have had a finite beginning
---the big-bang--- at which not only matter but \emph{space-time
itself} is born.  If space is compact, matter \emph{as well as
space-time} end in the big-crunch singularity. In this respect,
general relativity took us back to St. Augustine's paradigm but in a
detailed, specific and mathematically precise form.  In semi-popular
articles and radio shows, relativists now like to emphasize that the
question ``what was there before?'' is rendered meaningless because
the notions of `before' requires a pre-existing space-time geometry.
We now have a new paradigm, a new dogma: In the Beginning there was
the Big Bang.

But as I pointed out in section \ref{s1}, general relativity is
incomplete and there is no reason to trust its predictions near
space-time singularities. We must fuse it with quantum physics and
let the new theory tell us what happens when matter and geometry
enter the Planck regime.

\subsection{Some key questions}
\label{s2.2}

If the smooth continuum of Minkowski and Einstein is only an
approximation, on the issue of the origin of the universe we are now
led to ask:

\begin{itemize}
\item How close to the big-bang does a smooth space-time of
    general relativity make sense?  Inflationary scenarios, for
    example, are based on a space-time continuum. Can one show
    from some first principles that this is a safe approximation
    already at the onset of inflation?

\item Is the big-bang singularity naturally resolved by quantum
    gravity? This possibility led to the development of the
    field of quantum cosmology in the late 1960s. The basic idea
    can be illustrated using an analogy to the theory of the
    hydrogen atom. In classical electrodynamics the ground state
    energy of this system is unbounded below. Quantum physics
    intervenes and, thanks to a non-zero Planck's constant, the
    ground state energy is lifted to a finite value,
    $-me^4/2\hbar^2 \approx - 13.6{\rm eV}$. Since it is the
    Heisenberg uncertainly principle that lies at the heart of
    this resolution and since the principle must feature also in
    quantum gravity, one is led to ask: Can a similar mechanism
    resolve the big-bang and big crunch singularities of general
    relativity?

\item Is a new principle/ boundary condition at the big bang or
    the big crunch essential? The most well known example of
    such a boundary condition is the `no boundary proposal' of
    Hartle and Hawking \cite{hh}. Or, do quantum Einstein
    equations suffice by themselves even at the classical
    singularities?

\item Do quantum dynamical equations remain well-behaved even at
    these singularities? If so, do they continue to provide a
    deterministic evolution? The idea that there was a
    pre-big-bang branch to our universe has been advocated in
    several approaches, most notably by the pre-big-bang
    scenario in string theory \cite{pbb} and ekpyrotic and
    cyclic models \cite{ekp1,ekp2} inspired by the brane world
    ideas. However, these are perturbative treatments which
    require a smooth continuum in the background. Therefore,
    their dynamical equations break down at the singularity
    whence, without additional input, the pre-big-bang branch is
    not joined to the current post-big-bang branch by a
    deterministic evolution. Can one improve on this situation?

\item  If there is a deterministic evolution, what is on the
    `other side'? Is there just a quantum foam from which the
    current post-big-bang branch is born, say a `Planck time
    after the putative big-bang'? Or, was there another
    classical universe as in the pre-big-bang and cyclic
    scenarios, joined to ours by deterministic equations?
\end{itemize}

Clearly, to answer such questions we cannot start by assuming that
there is a smooth space-time in the background. But already in the
classical theory, it took physicists several decades to truly
appreciate the dynamical nature of geometry and to learn to do
physics without recourse to a background. In quantum gravity, this
issue becomes even more vexing.%
\footnote{There is a significant body of literature on issue; see
e.g., \cite{as-book} and references therein. These difficulties are
now being discussed also in the string theory literature in the
context of the AdS/CFT conjecture.}

For simple systems, including Minkowskian field theories, the
Hamiltonian formulation generally serves as the royal road to
quantum theory. It was therefore adopted for quantum gravity by
Dirac, Bergmann, Wheeler and others. But absence of a background
metric implies that the Hamiltonian dynamics is generated by
constraints \cite{kk}. In the quantum theory, physical states are
solutions to quantum constraints. All of physics, including the
dynamical content of the theory, has to be extracted from these
solutions. But there is no external time to phrase questions about
evolution. Therefore we are led to ask:

\begin{itemize}

\item Can we extract, from the arguments of the wave function,
    one variable which can serve as \emph{emergent time} with
    respect to which the other arguments `evolve'? If not, how
    does one interpret the framework? What are the physical
    (i.e., Dirac) observables? In a pioneering work, DeWitt
    proposed that the determinant of the 3-metric can be used as
    an `internal' time \cite{bd}. Consequently, in much of the
    literature on the Wheeler-DeWitt (\WDW) approach to quantum
    cosmology, the scale factor is assumed to play the role of
    time, although sometimes only implicitly. However, in closed
    models the scale factor fails to be monotonic due to
    classical recollapse and cannot serve as a global time
    variable already in the classical theory. Are there better
    alternatives at least in the simple setting of quantum
    cosmology? If not, can we still make physical predictions?

\end{itemize}

Finally there is an ultraviolet-infrared tension.

\begin{itemize}

\item Can one construct a framework that cures the
    short-distance difficulties faced by the classical theory
    near singularities, while maintaining an agreement with it
    at large scales?

\end{itemize}

By their very construction, perturbative and effective
descriptions have no problem with the second requirement. However,
physically their implications can not be trusted at the Planck
scale and mathematically they generally fail to provide a
deterministic evolution across the putative singularity. Since the
non-perturbative approaches often start from deeper ideas, it is
conceivable that they could lead to new structures at the Planck
scale which modify the classical dynamics and resolve the big-bang
singularity. But once unleashed, do these new quantum effects
naturally `turn-off' sufficiently fast, away from the Planck
regime? The universe has had some \emph{14 billion years} to
evolve since the putative big bang and even minutest quantum
corrections could accumulate over this huge time period leading to
observable departures from dynamics predicted by general
relativity. Thus, the challenge to quantum gravity theories is to
first create huge quantum effects that are capable of overwhelming
the extreme gravitational attraction produced by matter densities
of some $10^{105}\, {\rm gms/cc}$ near the big bang, and then
switching them off with extreme rapidity as the matter density
falls below this Planck scale. This is a huge burden!

These questions are not new; some of them were posed already in the
late sixties by quantum gravity pioneers such as Peter Bergmann,
Bryce DeWitt, Charles Misner and John Wheeler \cite{kk,bd,misner}.
However, the field reached an impasse in the late eighties.
Fortunately, this status-quo changed significantly over the last
decade with a dramatic inflow of new ideas from many directions. In
the next two subsections, I will summarize the current status of
these issues in loop quantum cosmology.

\subsection{FRW models and the \WDW theory}
\label{s2.3}

\begin{figure}[]
  \begin{center}
$a)$\hspace{8cm}$b)$
    \includegraphics[width=3.2in,angle=0]{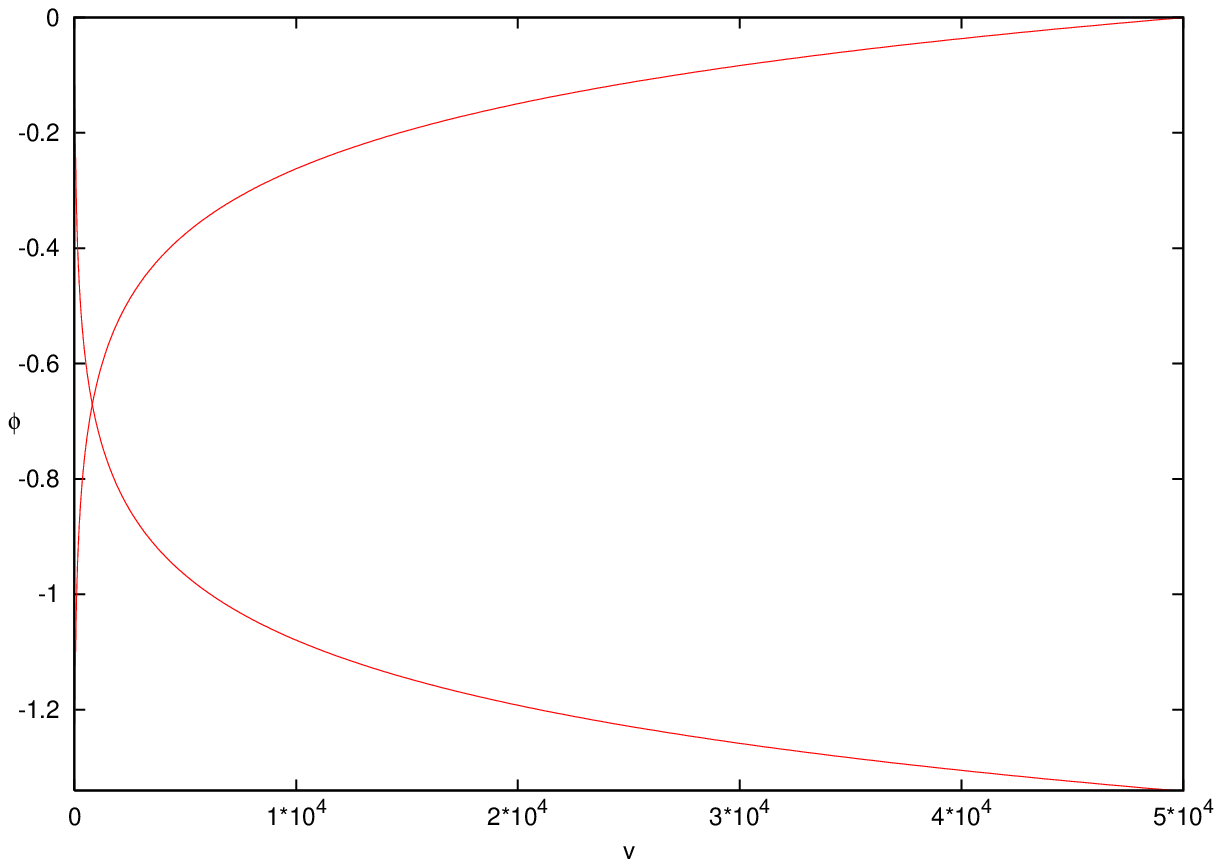}
\includegraphics[width=3.2in,angle=0]{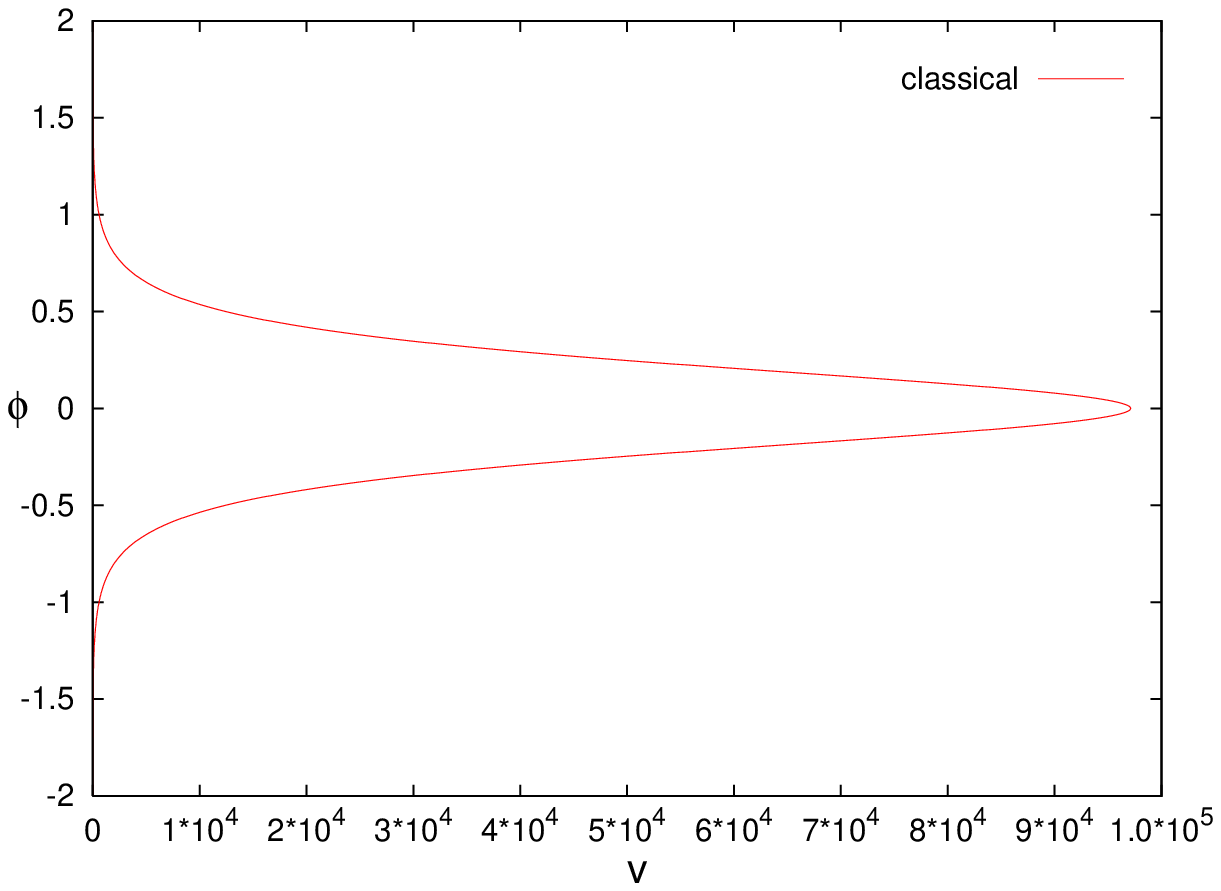}
\caption{$a)$ Classical solutions in k=0, $\Lambda=0$ FRW models with a
massless scalar field. Since $\pphi$ is a constant of motion,
a classical trajectory can be plotted in the $v$-$\phi$ plane,
where $v$ is the volume (essentially in Planck units). There
are two classes of trajectories. In one
the universe begins with a big-bang and expands and in the
other it contracts into a big crunch.
$b)$ Classical solutions in the k=1, $\Lambda =0$ FRW model with a massless
scalar field. The universe begins with a big bang, expands to a
maximum volume and then undergoes a recollapse to a big crunch
singularity. Since the volume is double valued in any solution,
it cannot serve as a global time coordinate in this case. The
scalar field on the other hand does so both in the k=0 and k=1
cases.}\label{class}
  \end{center}
\end{figure}
Almost all phenomenological work in cosmology is based on the k=0
homogeneous and isotropic Friedmann Robertson Walker (FRW)
space-times and perturbations thereof. For concreteness, I will
focus on FRW model in which the only matter source is a scalar
massless field.%
\footnote{Our discussion will make it clear that it is relatively
straightforward to allow additional fields, possibly with
complicated potentials.}
I will consider k=0 (or spatially flat) as well as k=1 (spatially
closed) models with or without a cosmological constant (of either
sign). Conceptually, these models are interesting for our purpose
because \emph{every} of their classical solutions has a singularity
(see Fig \ref{class}). Therefore a natural singularity resolution
without external inputs is highly non-trivial. In light of the
spectacular observational inputs over the past decade, the k=0 model
is the one that is phenomenologically most relevant. However as we
will see, because of its classical recollapse, the k=1 model offers
a more stringent viability test for the quantum cosmology.

In the classical theory, one considers one space-time at a time and
although the metric of that space-time is dynamical, it enables one
to introduce time coordinates that have direct physical
significance. However in the quantum theory ---and indeed already in
the phase space framework that serves as the stepping stone to
quantum theory--- we have to consider all possible homogeneous,
isotropic space-times. In this setting one can introduce a natural
foliation of the 4-manifold each leaf of which serves as the `home'
to a spatially homogeneous 3-geometry. However, unlike in
non-gravitational theories, there is no preferred physical
\emph{time variable} to define evolution. A natural strategy is to
use part of the system as an `internal' clock with respect to which
the rest of the system evolves. This leads one to Leibnitz's
\emph{relational time}. Now, in any spatially homogeneous model with
a massless scalar field $\phi$, the conjugate momentum $p_{(\phi)}$
is a constant of motion, whence $\phi$ is monotonic along any
dynamical trajectory. Thus, in the classical theory, it serves as a
global clock (see Fig \ref{class}). Questions about evolution can
thus be phrased as: ``If the curvature or matter density or an
anisotropy parameter is such and such when $\phi =\phi_1$ what is it
when $\phi = \phi_2$?'' What is the situation in the quantum theory?
There is no a priori guarantee that a variable which serves as a
viable time parameter in the classical theory will continue to do so
in the quantum theory. Whether it does so depends on the form of the
Hamiltonian constraint. For instance as Fig \ref{class}a shows, in
the k=0 model without a cosmological constant, volume (or the scale
factor) is a global clock along any classical trajectory. But the
form of the quantum Hamiltonian constraint \cite{aps2} in loop
quantum gravity is such that it does not serve this role in the
quantum theory. The scalar field, on the other hand, continues to do
so (also in the k=1 case and with or without a cosmological
constant).%
\footnote{If there is no massless scalar field, one could still use
a suitable matter field as a `local' internal clock. For instance in
the inflationary scenario, because of the presence of the potential
the inflaton is not monotonic even along classical trajectories. But
it is possible to divide dynamics into `epochs' and use the inflaton
as a clock locally, i.e., within each epoch \cite{aps4}. There is
considerable literature on the issue of internal time for model
constrained systems \cite{as-book} (such as a system of two harmonic
oscillators where the total energy is constrained to be constant
\cite{cr-time}).}

Because of the assumption of spatial homogeneity, in quantum
cosmology one has only a finite number of degrees of freedom.
Therefore, although the conceptual problems of quantum gravity
remain, there are no field theoretical infinities and one can hope
to mimic ordinary text book quantum mechanics to pass to quantum
theory.

However, in the k=0 case, because space is infinite, homogeneity
implies that the action, the symplectic structure and Hamiltonians
all diverge since they are represented as integrals over all of
space. Therefore, in any approach to quantum cosmology
---irrespective of whether it is based on path integrals or
canonical methods--- one has to introduce an elementary cell $\V$
and restrict all integrals to it. In actual calculations, it is
generally convenient also to introduce a fiducial 3-metric $\q_{ab}$
(as well as frames $\e^a_i$ adapted to the spatial isometries) and
represent the physical metric $q_{ab}$ via a scale factor $a$, \,
$q_{ab} = a^2\, \q_{ab}$. Then the geometrical dynamical variable
can be taken to be either $a$, or the oriented volume $v$ of the
fiducial cell $\V$ as measured by the physical frame $e^a_i$, where
$v$ is positive if $e^a_i$ has the same orientation as $\e^a_i$ and
negative if the orientations are opposite. (In either case the
physical volume of the cell is $|v|$.) In this chapter I will use
$v$ rather than the scale factor. Note, however, physical results
cannot depend on the choice of the fiducial $\V$ or $\q_{ab}$.%
\footnote{This may appear as an obvious requirement but
unfortunately it is often overlooked in the literature. The claimed
physical results often depend on the choice of $\V$ and/or $\q_{ab}$
although the dependence is often hidden by setting the volume $v_o$
of $\V$ with respect to $\q_{ab}$ to $1$ (in unspecified units) in
the classical theory.}
In the k=1 case, since space is compact, a fiducial cell is
unnecessary and the dynamical variable $v$ is then just the physical
volume of the universe.

With this caveat out of the way, one can proceed with quantization.
Situation in the \WDW theory can be summarized as follows. This
theory emerged in the late sixties and was analyzed extensively over
the next decade and a half \cite{kk}. Many of the key physical ideas
of quantum cosmology were introduced during this period
\cite{bd,misner} and a number of models were analyzed. However,
since a mathematically coherent approach to quantization of full
general relativity did not exist, there were no guiding principles
for the analysis of these simpler, symmetry reduced systems. Rather,
quantization was carried out following `obvious' methods from
ordinary quantum mechanics. Thus, in quantum kinematics, states were
represented by square integrable wave functions $\Psi(v,\phi)$,
where $v$ represents geometry and $\phi$, matter; and operators
$\hat{v}, \hat\phi$ acted by multiplication and their conjugate
momenta by ($-i \hbar$ times) differentiation. With these choices
The Hamiltonian constraint takes the form of a differential equation
that must be satisfied by the physical states\cite{aps3}:
\be \label{wdw0} \p_\phi^2 \ul\Psi(v,\phi) =
{\ul{\Theta}}_o\ul\Psi(v,\phi)\, := \, -12\pi G\, (v\partial_v)^2\,
\Psi(v,\phi)\, \ee
for k=0, and
\be \label{wdw1} \p_\phi^2 \ul\Psi(v,\phi) = -{\ul{\Theta}}_1
\ul\Psi(v,\phi) \, := \, -{\ul{\Theta}}_o \ul\Psi(v,\phi) -
G\,C\,|v|^{\frac{4}{3}}\, \ul\Psi(v,\phi) \ , \ee
for k=1, where $C$ is a numerical constant. \emph{In what follows
$\ul{\Theta}$ will stand for either $\ul{\Theta}_o$ or
$\ul{\Theta}_1$.} In the older literature, the emphasis was on
finding and interpreting the WKB solutions of these equations (see,
e.g., \cite{ck}). However, near the singularity, the WKB
approximation fails and we need an exact quantum theory.

The exact theory can be readily constructed \cite{aps2,aps3}. Note
first that the form of (\ref{wdw0}) and (\ref{wdw1}) is the same as
that of a Klein-Gordon equation in a 2-dimensional static space-time
(with a $\phi$-independent potential in the k=1 case), where $\phi$
plays the role of time and $v$ of space. This suggests that we think
of $\phi$ as the relational time variable with respect to which $v$,
the `true' degree of freedom, evolves. A systematic procedure based
on the so-called group averaging method \cite{dm} (which is
applicable for a very large class of constrained systems) then leads
us to the physical inner product between these states. Not
surprisingly it coincides with the expression from the Klein-Gordon
theory in static space-times.

\begin{figure}[]
  \begin{center}
    \includegraphics[width=3.2in,angle=0]{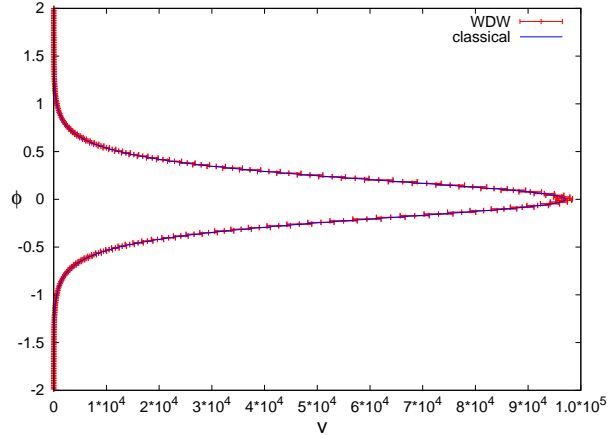}
\caption{ Expectation values (and dispersions) of ${|\hat{v}|_{\phi}}$
for the WDW wave function in the k=1 model. The \WDW wave function
follows the classical trajectory into the big-bang and big-crunch
singularities. (In this simulation, the parameters were:
$p_{\phi}^{\star} = 5000 $, and $\Delta p_\phi/p_\phi^{\star} = 0.02$.) }
    \label{fig:wdw}
 \end{center}
\end{figure}

The physical sector of the final theory can be summarized as
follows. The physical Hilbert space $\Hp$ in the k=0 and k=1 cases
consists of `positive frequency' solutions to (\ref{wdw0}) and
(\ref{wdw1}) respectively. A complete set of observables is provided
by the momentum $\hat\pphi$ and the relational observable
$\hat{|v|}|_\phi$ representing the volume at the `instant of time
$\phi$':
\be \h\pphi = -i\hbar \partial_\phi\quad{\rm and}\quad \h{V}|_\phi =
e^{i\ul{\Theta}\,(\phi-\phi_o)}\, |v|\, e^{- i\ul {\Theta}
\,(\phi-\phi_o)}\ee
There are \emph{Dirac} observables because their action preserves
the space of solutions to the constraints and are self-adjoint on
the physical Hilbert space $\Hp$. With the exact quantum theory at
hand, we can ask if the singularities are naturally resolved. More
precisely, from $\h\pphi$ and $\h{V}|_\phi$ we can construct
observables corresponding to matter density $\h\rho$ (or space-time
scalar curvature $\h{R}$). Since the singularity is characterized by
divergence of these quantities in the classical theory, in the
quantum theory we can proceed as follows. We can select a point
$(v_o, \phi_o)$ at a `late time' $\phi_o$ on a classical trajectory
of Fig \ref{class} ---e.g., now, in he history of our universe---
when the density and curvature are \emph{very} low compared to the
Planck scale, and construct a semi-classical state which is sharply
peaked at $v_o$ at $\phi =\phi_o$. We can then evolve this state
\emph{backward} in time. Does it follow the classical trajectory? To
have the correct `infra-red' behavior, it must, until the density
and curvature become very high. What happens in this `ultra-violet'
regime? Does the quantum state remain semi-classical and follow the
classical trajectory into the big bang? Or, does it spread out
making quantum fluctuations so large that although the quantum
evolution does not break down, there is no reasonable notion of
classical geometry? Or, does it remain peaked on some trajectory
which however is so different from the classical one that, in this
backward evolution, the the universe `bounces' rather than being
crushed into the singularity? Or, does it ... Each of these
scenarios provides a distinct prediction for the ultra-violet
behavior and therefore for physics in the deep Planck regime.%
\footnote{ Sometimes apparently weaker notions of singularity
resolution are discussed. Consider two examples \cite{kks}. One may
be able to show that the wave function vanishes at points of the
classically singular regions of the configuration space. However, if
the \emph{physical} inner product is non-local in this configuration
space ---as the group averaging procedure often implies--- such a
behavior of the wave function would not imply that the probability
of finding the universe at these configurations is zero. The second
example is that the wave function may become highly non-classical.
This by itself would not mean that the singularity is avoided unless
one can show that the expectation values of a family of Dirac
observables which become classically singular remain finite there.}

It turns out that the \WDW theory leads to similar predictions in
both k=0 and k=1 cases \cite{aps2,aps3,apsv}. They pass the
infra-red tests with flying colors (see Fig \ref{fig:wdw}). But
unfortunately the state follows the classical trajectory into the
big bang (and in the k=1 case also the big crunch) singularity. Thus
the first of the possibilities listed above is realized. The
singularity is not resolved because expectation values of density
and curvature continue to diverge in epochs when their classical
counterparts do. The analogy to the hydrogen atom discussed in
section \ref{s2.2} fails to be realized.

\subsection{Loop quantum cosmology: New quantum mechanics}
\label{s2.4}

For a number of years, the failure of the \WDW theory to naturally
resolve the big bang singularity was taken to mean that quantum
cosmology cannot, by itself, shed any light on the quantum nature of
the big bang. Indeed, for systems with a finite number of degrees of
freedom we have the von Neumann uniqueness theorem which guarantees
that quantum kinematics is unique. The only freedom we have is in
factor ordering and this was deemed insufficient to alter the
status-quo provided by the \WDW theory.

The situation changed dramatically in LQG. Here, a well established,
rigorous kinematical framework \emph{is} available for full general
relativity \cite{almmt,alrev,crbook,ttbook}. If one mimics it in
symmetry reduced models, one is led to a quantum theory which is
\emph{inequivalent to that of the \WDW theory already at the
kinematic level}. Quantum dynamics built in this new arena agrees
with the \WDW theory in `tame' situations but differs dramatically
in the Planck regime, leading to a natural resolution of the big
bang singularity.

But what about the von Neumann uniqueness theorem? The theorem
states that 1-parameter groups $U(\lambda)$ and $V(\mu)$
satisfying the Weyl commutation relations%
\footnote{These are: $U(\lambda V(\lambda) = e^{i\lambda\mu} V(\mu)
U(\lambda)$ and can be obtained by setting $U(\lambda) = e^{i\lambda
\h{x}}$ and $V(\mu) = e^{i\mu \h{p}}$ in the standard Schr\"odinger
theory. Given a representation $U(\lambda)$ is said to be
\emph{weakly continuous} in $\lambda$ if its matrix elements between
any two fixed quantum states are continuous in $\lambda$.}
admit (up to isomorphism) a unique irreducible representation by
unitary operators on a Hilbert space $\H$ in which $U(\lambda)$ and
$V(\mu)$ are weakly continuous in the parameters $\lambda$ and
$\mu$. By Stone's theorem, weak continuity is the necessary and
sufficient condition for $\H$ to admit self adjoint operators
$\h{x}, \h{p}$ such that $U(\lambda) = e^{i\lambda \h{x}}$ and
$V(\mu) = e^{i\mu \h{p}}$. Therefore assumption of the von Neumann
theorem are natural in non-relativistic quantum mechanics and we are
led to a unique quantum kinematics. However, in full loop quantum
gravity, $x$ is analogous to the gravitational connection and
$U(\lambda)$ to its holonomy. One can again construct an abstract
algebra using holonomies and operators conjugate to connections and
ask for its representations satisfying natural assumptions the most
important of which is the diffeomorphism invariance dictated by
background independence. There is again a uniqueness theorem
\cite{lost}. However, in the representation that is thus singled
out, holonomy operators ---analogs of $U(\lambda)$--- fails to be
weakly continuous whence there are no operators corresponding to
connections! Furthermore, a number of key features of the theory
---such as the emergence of a quantum Riemannian geometry in which
there is fundamental discreteness--- can be traced back to this
unforeseen feature. Therefore, upon symmetry reduction, although we
have a finite number of degrees of freedom, it would be incorrect to
just mimic Schr\"oddinger quantum mechanics and impose weak
continuity. When this assumption is dropped, the von Neumann theorem
is no longer applicable and \emph{we have new quantum mechanics}
\cite{abl}.

Thus, the key difference between LQC and the \WDW theory lies in the
fact that while one does not have reliable quantum kinematics in the
\WDW theory, there is a well developed and rigorous framework in LQG
which, furthermore, is \emph{unique}! If we mimic it as closely as
possible in the symmetry reduced theories, we are led to a new
kinematic arena, distinct from the one used in the \WDW quantum
cosmology. LQC is based on this arena.

\subsection{LQC: Dynamics}
\label{s2.5}

It turns out \WDW dynamics is not supported by the new arena
because, when translated in terms of gravitational connections and
their conjugate momenta, it requires that there be an operator
corresponding to the connection itself. Therefore one has to develop
quantum dynamics ab-initio on the new arena. The result is that the
differential operator $\ul\Theta_o\, =\, -12\pi G\,
(v\partial_v)^2\, $ in Eqs (\ref{wdw0}) and (\ref{wdw1}) is now
replaced by a second order \emph{difference} operator in $v$, where
the step size is dictated by the `area gap' of LQG, i.e., the lowest
non-zero eigenvalue of the area operator in LQG. There is a precise
sense in which the Wheeler-DeWitt equations result as the limits of
LQC equations when the area gap is taken to zero, i.e., when the
Planck scale discreteness of quantum geometry determined by LQG is
neglected. We will now see that this discreteness is completely
negligible at late times but plays a crucial role in the Planck
scale geometry near singularities.

The LQC dynamics has been analyzed using three different methods.
\begin{itemize}

\item Numerical solutions of the exact quantum equations
    \cite{aps1,aps2,aps3,apsv}. A great deal of effort was spent in
    ensuring that the results are free of artifacts of simulations,
    do not depend on the details of how semi-classical states are
    constructed and hold for a wide range of
    parameters.\\
\item Effective equations \cite{jw,aps3,apsv}. These are
    differential equations which include the leading quantum
    corrections. The asymptotic series from which these
    contributions were picked was constructed rigorously but is
    based on assumptions whose validity has not been established.
    Nonetheless the effective equations approximate the exact
    numerical evolution of semi-classical states extremely well.\\
\item Exactly soluble, but simplified model in the k=0 case
    \cite{mb-exact,acs}. The simplification is well controlled
    \cite{acs}. This analysis has provided some results which
    provide an analytical understanding of numerical results and
    also several other results which are not restricted to states
    which are semi-classical at late times. In this sense the
    analysis shows that the overall picture is robust within these
    models.
\end{itemize}

I will provide a global picture that has emerged from these
investigations, first for the k=1 model without the cosmological
constant $\Lambda$ and for the k=0 case for various values of
$\Lambda$.

Recall that in classical general relativity, the k=1 closed
universes start out with a big bang, expand to a maximum volume
$V_{\rm max}$ and then recollapse to a big-crunch singularity.
Consider a classical solution in which $V_{\rm max}$ is
astronomically large ---i.e., on which the constant of motion
$p_{(\phi)}$ takes a large value $p_{(\phi)}^\star$--- and consider
a time $\phi_o$ at which the volume $v^\star$ of the universe is
also large. Then there are well-defined procedures to construct
states $\Psi(v,\phi)$ in the \emph{physical Hilbert space} which are
sharply peaked at these values of observables $\h{p}_{(\phi)}$ and
$\h{V}_{\phi_o}$ at the `time' $\phi_o$. Thus, at `time' $\phi_o$,
the quantum universe is well approximated by the classical one. What
happens to such quantum states under evolution? As emphasized
earlier, there are infra-red and ultra-violet challenges:\\
i) Does the state remain peaked on the classical trajectory in the
low curvature regime? Or, do quantum geometry effects accumulate
over the cosmological time scales, causing noticeable deviations
from classical general relativity? In particular, is there a
recollapse and if so does the value $V_{\rm max}$ of maximum volume
agree with that predicted by general relativity \cite{gu}?\\
ii) What is the behavior of the quantum state in the Planck regime?
Is the big-bang singularity resolved? What about the big-crunch? If
they are both resolved, what is on the `other side'?

\begin{figure}[]
  \begin{center}
    $a)$\hspace{8cm}$b)$
    \includegraphics[width=3.2in,angle=0]{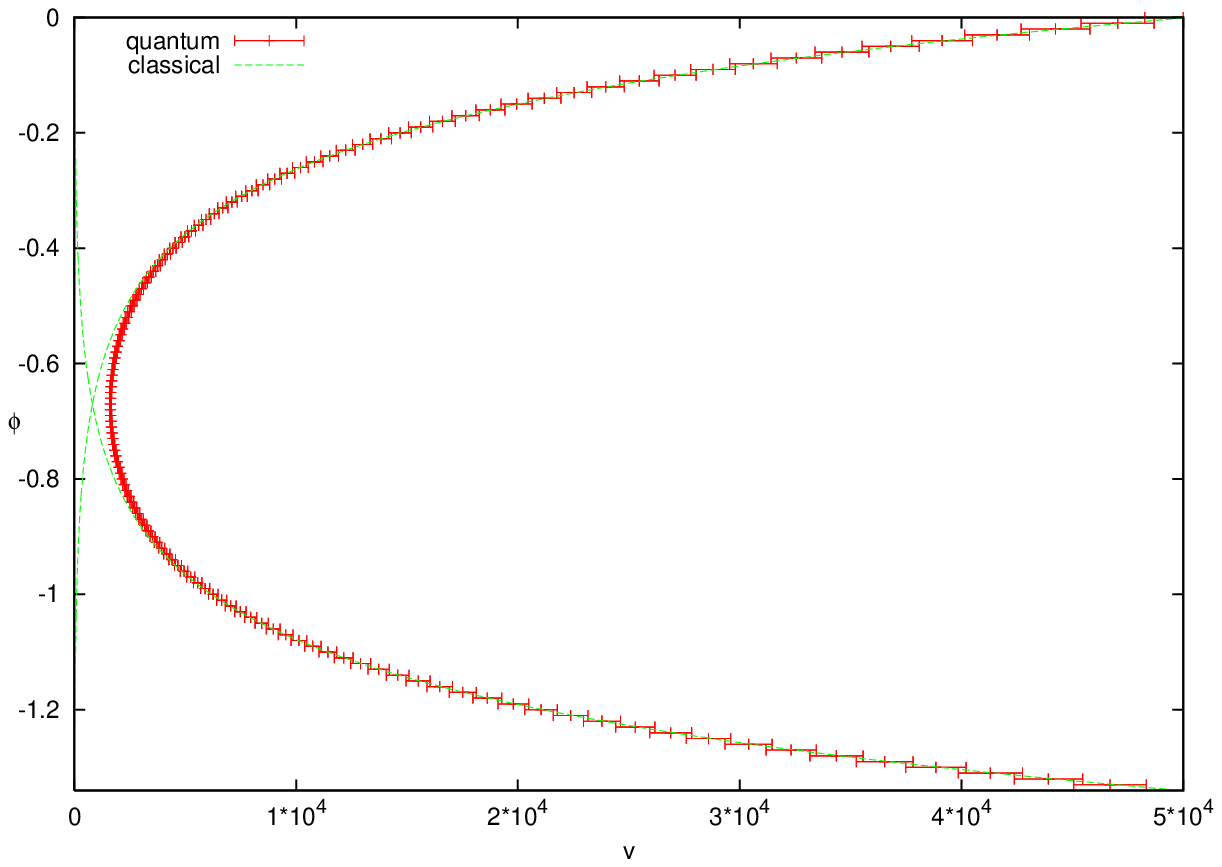}
    \includegraphics[width=3.2in,angle=0]{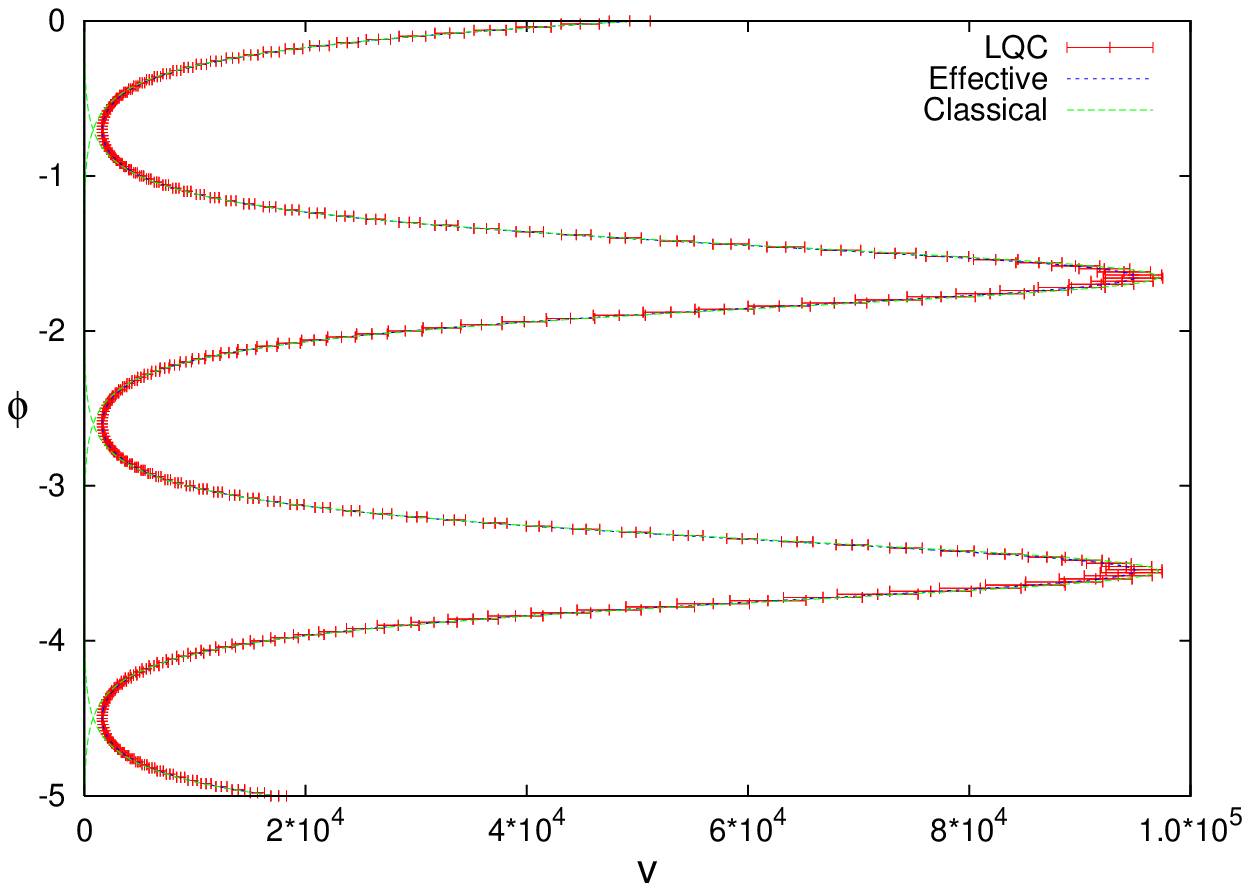}
\caption{In the LQC evolution of models under consideration, the big
bang and big crunch singularities are replaced by quantum bounces.
Expectation values and dispersion of $|\h{v}|_\phi$, are compared
with the classical trajectory and the trajectory from effective
Friedmann dynamics. The classical trajectory deviates significantly
from the quantum evolution at Planck scale and evolves into
singularities. The effective trajectory provides an excellent
approximation to quantum evolution at all scales. \,\, $a)$ The k=0
case. In the backward evolution, the quantum evolution follows our
post big-bang branch at low densities and curvatures but undergoes a
quantum bounce at matter density $\rho \sim 0.82\rho_{\rm PL}$ and
joins on to the classical trajectory that was contracting to the
future. $b)$ The k=1 case. The quantum bounce occurs again at $\rho \sim 0.82
\rho_{\rm Pl}$. Since the big bang and the big crunch singularities
are resolved the evolution undergoes cycles. In this simulation
$p_{(\phi)}^\star = 5\times 10^3$, $\Delta
p_{(\phi)}/p_{(\phi)}^\star = 0.018$, and $v^\star = 5\times 10^4$.}
\label{fig:lqc}
\end{center}
\end{figure}

Numerical simulations show that the wave functions do remain sharply
peaked on classical trajectories in the low curvature region also in
LQC. But there is a radical departure from the \WDW results in the
strong curvature region. The \WDW evolution follows classical
dynamics all the way into the big-bang and big crunch singularities
(see Fig \ref{fig:wdw}). In LQC, by contrast, \emph{the big bang and
the big crunch singularities are resolved and replaced by
big-bounces} (see Fig \ref{fig:lqc}). In these calculations, the
required notion of semi-classicality turns out to be surprisingly
weak: these results hold even for universes with $a_{\rm max}
\approx 23 \lp$ and the `sharply peaked' property improves greatly
as $a_{\mathrm{max}}$ grows.

More precisely, numerical solutions have shown that the situation is
as follows. (For details, see \cite{apsv}.)\smallskip

\b The trajectory defined by the expectation values of the physical
observable $\h{V}|_\phi$ in the full quantum theory is in good
agreement with the trajectory defined by the classical Friedmann
dynamics until the energy density $\rho$ in the matter field is
about two percent of the Planck density. In the classical solution,
scalar curvature and the matter energy density keep increasing on
further evolution, eventually leading to a big bang (respectively,
big crunch) singularity in the backward (respectively, forward)
evolution, where $v \rightarrow 0$. The situation is very different
with quantum evolution. As the density and curvature increases
further, quantum geometry effects become dominant creating an
effective repulsive force which rises very quickly, overwhelms
classical gravitational attraction, and causes a bounce at $\rho
\sim 0.82\rho_{\rm Pl}$, thereby resolving the past (or the big
bang) and future (or the big crunch) singularities. There is thus a
cyclic scenario depicted in Fig \ref{fig:lqc}.

\b The volume of the universe takes its minimum value $V_{\rm min}$
at the bounce point. $V_{\rm min}$ scales linearly with
$p_{(\phi)}$:
\footnote{Here and in what follows, numerical values are given in
the classical units $G=c=1$. In these units $p_{(\phi)}$ has the same
physical dimensions as $\hbar$ and the numerical value of $\hbar$ is
$2.5\times 10^{-66}{\cm}^2. $}
\be V_{\rm min} = \big( \f{4\pi G\gamma^2 \Delta}{3}
\big)^{\f{1}{2}}\, p_{(\phi)} \,\, \approx (1.28\times 10^{-33}\,
{\cm})\,\, p_{(\phi)}\ee
Consequently, $V_{\rm min}$ can be \emph{much} larger than the
Planck size.  Consider for example a quantum state describing a
universe which attains a maximum radius of a megaparsec. Then the
quantum bounce occurs when the volume reaches the value $V_{\rm min}
\approx 5.7 \times 10^{16}\, {\cm}^3$, \emph{some $10^{115}$ times
the Planck volume.} Deviations from the classical behavior are
triggered when the density or curvature reaches the Planck scale.
The volume can be very large and is not the relevant scale for
quantum gravity effects.

\b After the quantum bounce the energy density of the universe
decreases and the repulsive force dies quickly when matter density
reduces to about two percept of the Planck density. The quantum
evolution is then well-approximated by the classical trajectory. On
subsequent evolution, the universe recollapses both in classical and
quantum theory at the value $V=V_{\mathrm{max}}$ when energy density
reaches a minimum value $\rmin$.  $V_{\rm max}$ scales as the
3/2-power of $p_{(\phi)}$:
\be\label{Vmax}  V_{\rm max} = (16\pi G/3 \lo^2)^{{3/4}}\,
p_{(\phi)}^{{3/2}}\,\, \approx \,\,  0.6\,
p_{(\phi)}^{{3/2}} \ee
Quantum corrections to the classical Friedmann formula $\rmin =
3/8\pi Ga^2_{\rm max}$ are of the order $O(\lp/a_{\rm max})^4$. For
a universe with $a_{\rm max} = 23\lp$, the correction is only one
part in $10^{5}$. For universes which grow to macroscopic sizes,
classical general relativity is essentially exact near the
recollapse.

\b Using ideas from geometrical quantum mechanics \cite{as}, one can
obtain certain effective classical equations which incorporate the
leading quantum corrections \cite{jw,apsv}.  While the classical
Friedmann equation is $(\dot{a}/a)^2 = ({8\pi G}/{3})\, (\rho-
3/8\pi G a^2)$, the effective equation turns out to be
\be \label{eff} \left(\f{\dot{a}}{a} \right)^2 = \fs{8\pi G}{3}\,
(\rho-\rho_1(v))\,\, \left[f(v) - \f{\rho}{\rcr} \right] \ee
where $\rho_1$ and $f$ are specific functions of $v$ with $\rho_1
\sim 3/8\pi G a^2$. Bounces occur when $\dot{a}$ vanishes, i.e. at
the value of $v$ at which the matter density equals $\rho_1(v)$ or
$f(v) = \rho/\rcr$. The first root $\rho(v) = \rho_1(v)$ corresponds
to the classical recollapse while the second root, $f(v)=
\rho/\rcr$, to the quantum bounce. Away from the Planck regime, $f
\approx 1$ and $\rho/\rho_{\rm crit} \approx 0$. Bounces occur when
$\dot{a}$ vanishes, i.e. at the value of $v$ at which the matter
density equals $\rho_1(v)$ or $\rho_2(v)$.

\b For quantum states under discussion, the density $\rmax$ is well
approximated by $\rcr \approx 0.82 \rho_{\mathrm{Pl}}$ up to terms
$O(\lp^2/a_{\rm min}^2)$, independently of the details of the state
and values of $p_{(\phi)}$. (For a universe with maximum radius of a
megaparsec, $\lp^2/a_{\rm min}^2 \approx 10^{-76}$.) The density
$\rmin$ at the recollapse point also agrees with the value $(3/8 \pi
G a^2_{\rm max})$ predicted by the classical evolution to terms of
the order $O(\lp^4/a_{\rm max}^4)$. Furthermore the scale factor
$a_{\rm max}$ at which recollapse occurs in the quantum theory
agrees to a very good precision with the one predicted by the
classical dynamics.

\b The trajectory obtained from effective Friedmann dynamics is in
excellent agreement with quantum dynamics \emph{throughout the
evolution.}  In particular, the maximum and the minimum energy
densities predicted by the effective description agree with the
corresponding expectation values of the density operator $\hat \rho
\equiv \widehat{p_{(\phi)}^2/2|p|^3}$ computed numerically.

\b The state remains sharply peaked for a \emph{very large number of
`cycles'.} Consider the example of a semi-classical state with an
almost equal relative dispersion in $p_{(\phi)}$ and $|v|_\phi$ and
peaked at a large classical universe of the size of a megaparsec.
When evolved, it remains sharply peaked with relative dispersion in
$|v|_\phi$ of the order of $10^{-6}$ \emph{even after $10^{50}$
cycles of contraction and expansion!} Any given quantum state
eventually ceases to be sharply peaked in $|v|_\phi$ (although it
continues to be sharply peaked in the constant of motion
$p_{(\phi)}$). Nonetheless, the quantum evolution continues to be
deterministic and well-defined for an infinite number of cycles.
This is in sharp contrast with the classical theory where the
equations break down at singularities and there is
no deterministic evolution from one cycle to the next. \\

This concludes the summary of our discussion of the k=1 model.  An
analogous detailed analysis has been carried out also in the k=0
model, again with a free massless scalar field
\cite{aps1,aps2,aps3,acs}. In this case, if the cosmological
constant $\Lambda$ vanishes, as Fig \ref{class} shows, classical
solutions are of two types, those which start out at the big-bang
and expand out to infinity and those which start out with large
volume and contract to the big crunch singularity. Again, in this
case while the \WDW solution follows the classical trajectories into
singularities, the LQC solutions exhibit a big bounce. The LQC
dynamics is again faithfully reproduced by an effective equation:
the Friedmann equation $(\dot{a}/a)^2 = (8\pi G\, \rho/3)$ is
replaced just by $(\dot{a}/a)^2 = (8\pi G\,\rho /3)\, (1 -
\rho/\rcr)$. The quantum correction $\rho/\rcr$ is completely
negligible even at the onset of the standard inflationary era.
Quantum bounce occurs at $\rho=\rcr$ and the critical density is
again given by $\rcr \approx 0.82\rho_{\rm Pl}$. Furthermore, one
can show that the spectrum of the density operator \emph{on the
physical Hilbert space} admits a finite upper bound $\rho_{\rm
sup}$. By plugging values of constants in the analytical expression
of this bound, one finds $\rho_{\rm sup} = \rcr$! If $\Lambda >0$,
there are again two types of classical trajectories but the one
which starts out at the big-bang expands to an infinite volume in
finite value $\phi_{\rm max}$ of $\phi$. (The other trajectory is a
`time reverse' of this.) Because the $\phi$ `evolution' is unitary
in LQC, it yields a natural extension of the classical solution
beyond $\phi_{\rm max}$. If $\Lambda <0$, the classical universe
undergoes a recollapse. This is faithfully reproduced by the LQC
evolution. Since both the big-bang and the big-crunch singularities
are resolved, the LQC evolution leads to a cyclic universe as in the
k=1 model. Thus, in all these cases, the principal features of the
LQC evolution are robust, including the value of $\rcr$.

Let us summarize the overall situation. In simple cosmological
models, all the questions raised in section \ref{s2.2} have been
answered in LQC in remarkable detail. The scalar field plays the
role of an internal or emergent time and enables us to interpret the
Hamiltonian constraint as an evolution equation. The matter momentum
$\h\pphi$ and `instantaneous' volumes $\h{V}|_\phi$ form a complete
set of Dirac observables and enable us to ask physically interesting
questions. Answers to these questions imply that the big bang and
the big crunch singularities are naturally replaced by quantum
bounces. On the `other side' of the bounce there is again a large
universe. General relativity is an excellent approximation to
quantum dynamics once the matter density falls below a couple of
percent of the Planck density. Thus, LQC successfully meets both the
`ultra-violet' and `infra-red' challenges. Furthermore results
obtained in a number of models using distinct methods re-enforce one
another. One is therefore led to take at least the qualitative
findings seriously: \emph{Big bang is not the Beginning nor the big
crunch the End.} Quantum space-time appears to be vastly larger than
what general relativity had us believe!

\section{Black Holes}
\label{s3}

The idea of black holes is quite old. Already in 1784, in an
article in  the Proceedings of the Royal Society John Mitchell
used the formula for escape velocity in Newtonian gravity to argue
that light can not escape from a body of mass $M$ if it is
compressed to a radius $R = 2GM/c^2$. He went on to say
\begin{quote}
\emph{if there should exist in nature any [such] bodies .... we
could have no information from sight; yet if any other luminous
bodies should happen to revolve around them we might still perhaps
from the motions of these revolving bodies infer the existence of
the central ones with some degree of probability.}
\end{quote}
Remarkably, it is precisely observations of this type that have now
led us to the conclusion that there is a 3.4 million solar mass
black hole in the center of our galaxy! In the second volume of
Exposition du syst\`em du Monde published in 1798, the Marquis de
Laplace came to the same conclusion independently and was more
confident of the existence of black holes:
\begin{quote}
\emph{there exist, in the immensity of space, opaque bodies as
considerable in magnitude, and perhaps equally as numerous as
stars.}
\end{quote}

While in many ways these observations are astonishingly prescient,
the underlying reasoning is in fact incorrect. For, if light (which
is assumed to be corpuscular in this argument) from a distant source
were to impinge on such an object, it would bounce back and by
Newtonian conservation laws it would reach the point from which it
came. Distant observers should therefore be able to see these
objects. Indeed, if all speeds ---including that of light--- are
relative as in Newtonian mechanics, there can really be no black
holes. The existence of black holes requires both gravity and an
absolute speed of light; general relativity is essential.

\subsection{Horizons}
\label{s3.1}

To capture the intuitive notion that black hole is a region from
which signals can not escape to the asymptotic part of space-time,
one needs a precise definition of future infinity. The standard
strategy is to use Penrose's conformal boundary $\sp$ \cite{rp}. It
is a future boundary: No point of the physical space-time lies to
the future of any point of $\sp$. It has topology $\S^2\times \R$
and it is null (assuming that the cosmological constant is zero). In
Minkowski space-time, one can think of $\sp$ as the `final resting
place' of all future directed null geodesics. More precisely, the
chronological past $I^-(\sp)$ of $\sp$ is
entire Minkowski space.%
\footnote{$I^-(\sp)$ is the set of all points in the physical
space-time from which there is a future directed time-like curve to
a point on $\sp$ in the conformally completed space-time. The term
`chronological' refers to the use of time-like curves. A curve which
is everywhere time-like or null is called `causal'.}

Given a general asymptotically flat space-time $(M,g_{ab})$, one
first finds the chronological past $I^-(\sp)$ of $\sp$. If it is not
the entire space-time, then there is a region in $(M,g_{ab})$ from
which one cannot send causal signals to infinity. When this happens,
one says that the space-time admits a black hole. More precisely,
\emph{Black-hole region} $\B$ of $(M,g_{ab})$ is defined as
\be \B = M \,-\, I^{-}(\sp)\,  \ee
where the right side is the set of points of $M$ which are not in
$I^{-}(\sp)$. The boundary $\partial\B$ of the black hole region is
called the \emph{event horizon} (EH) and is denoted by $E$
\cite{swh2}. $I^-(\sp)$ is often referred to as the asymptotic
region and $\E$ is the boundary of this region within physical
space-time.

Event horizons and their properties have provided a precise arena to
describe black holes and their dynamics. In particular, we have the
celebrated result of Hawking's \cite{swh1,swh2}: assuming energy
conditions, the area $a_{\hor}$ of an EH cannot decrease under time
evolution. The area $a_{\hor}$ is thus analogous to thermodynamic
entropy. There are other laws governing black holes which are in
equilibrium (i.e. stationary) and that make transitions to nearby
equilibrium states due to influx of energy and angular momentum.
They are similar to the zeroth and the first law of thermodynamics
and suggest that the surface gravity $\kappa$ of stationary black
holes is the analog of thermodynamic temperature. These analogies
were made quantitative and precise by an even deeper result Hawking
obtained using quantum field theory in a black hole background
\cite{swh3}: black holes radiate quantum mechanically as though they
are black bodies at temperature $T= \kappa\hbar/2\pi$. Their entropy
is then given by $S = a_{\hor}/4\lp^2$. Not surprisingly these
results have led to a rich set of insights and challenges over the
last 35 years.

\begin{figure}[]
  \begin{center}
    \includegraphics[width=1.8in,angle=0]{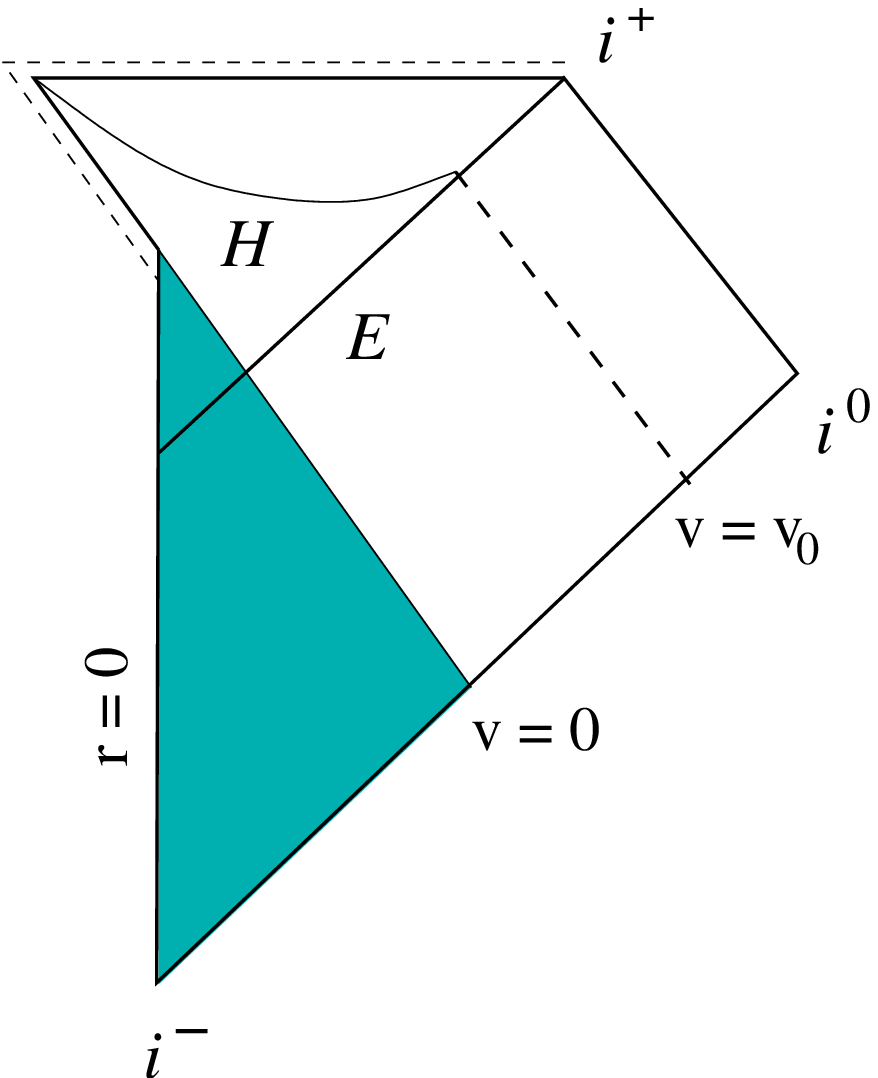}
     \hskip3.3cm
    \includegraphics[width=1.4in,angle=0]{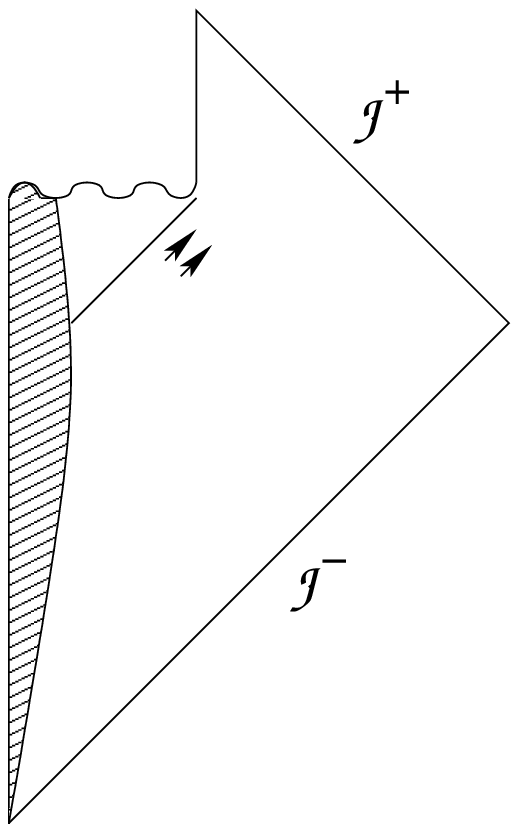}
\caption{$a)$ A Vaidya solution: Collapse of a spherical null
fluid to form a black hole. The null fluid radiation starts at the
retarded time $v=0$ and ends at $v=v_0$. Space time is flat in the
past of $v=0$ and Schwarzschild to the future of $v=v_0$. The
dynamical horizon $H$ starts out space-like and joins on to the
null event horizon at $v=v_0$. The event horizon first forms and
grows in the flat part of space-time. $b)$ Conjectured Penrose
diagram of an evaporating black hole: A black hole forms by
stellar collapse and evaporates due to Hawking radiation. Due to
back reaction, the singularity loses its strength as we move right
along the wiggly line, and finally disappears. Nonetheless because
there is still a piece of space-like singularity in the future,
$\mathcal{I}^+$ does not constitute the full future boundary of
space-time, leading to information loss.} \label{fig:collapse}
\end{center}
\end{figure}

However, the notion of an EH also has two severe limitations. First,
while the notion neatly captures the idea that asymptotic observers
can not `look into' a black hole, it is too global for many
applications. For example, since it refers to null infinity, it can
not be used in spatially compact space-times. Asymptotic flatness
and the notion of $\sp$ is used also in other contexts, in
particular to discuss gravitational radiation in full, non-linear
general relativity \cite{rp}. However, there $\sp$ is used just to
facilitate the imposition of boundary condition and make notions
such as `$1/r^n$-fall-off' precise. Situation with EHs is quite
different because they refer to the \emph{full chronological past
of} $\sp$. As a consequence, by changing the geometry in a small
---say Planck scale region--- around the singularity, one can change
the EH dramatically and even make it disappear \cite{ph}! As I
explained in section \ref{s1}, there is no reason to trust classical
general relativity very close to the singularity. If the singularity
is resolved due to quantum effects, there may be no longer an EH.
What then is a black hole? If the notion continues to be meaningful,
can we still associate with it entropy in absence of an EHs?

The second limitation is that the notion is teleological; it lets us
speak of a black hole \emph{only after we have constructed the
entire space-time}. Thus, for example, an EH may well be developing
in the room you are now sitting \emph{in anticipation} of a
gravitational collapse that may occur in this region of our galaxy a
million years from now. Indeed, as Fig \ref{fig:collapse}a shows,
EHs can form and grow \emph{even in flat space-time} where there is
no influx of matter of radiation. How can we then attribute direct
physical significance to the growth of their area? Clearly, when
astrophysicists say that they have discovered a black hole in the
center of our galaxy, they are referring to something much more
concrete and quasi-local than an EH.

Over the last five years, quasi-local horizons were introduced to
improve on this situation \cite{ihprl,ak1,ak2,akrev}. The idea is to
use the notion of marginally trapped surfaces. Consider a space-like
2-sphere in Minkowski space and illuminate it instantaneously. Then
there are two light fronts, one traveling outside the sphere and
expanding continuously and the other traveling inside and
contracting. Now, if the 2-sphere were placed in a strong
gravitational field, both these light fronts could contract. Then
light would be trapped and the sphere would not be visible from
outside. These two situations are separated by the marginal case
where one light front would be contracting and the area of the other
would neither decrease not increase. Such 2-surfaces are said to be
\emph{marginally trapped} and their world tubes represent
quasi-local horizons. More precisely, a marginally trapped tube
(MTT) is a 3-manifold which is foliated by a family of marginally
trapped 2-spheres. If it is space-like, the area of the marginally
trapped surfaces increases to the future and the MTT is called a
\emph{dynamical horizon} (DH). Heuristically it represents a growing
black hole. If the MTT is null, it is called an \emph{isolated
horizon} (IH) and represents a black hole in equilibrium. In Fig
\ref{fig:collapse}a a DH $H$ forms due to gravitational collapse of
infalling null fluid, grows in area with the in-fall and settles
down to an IH which coincides with the future part of the EH $E$.
Note that the definitions of MTT, DH and IH are all quasi-local. In
particular, they are not teleological; you can be rest assured that
none of these quasi-local horizons exists in the room you are now
sitting in!

There is however a significant drawback: lack of uniqueness.
Although partial uniqueness results exist \cite{ag}, in general we
cannot yet associate a unique DH with a generic, growing black hole.
But this weakness is compensated in large measure by the fact that
interesting results hold for \emph{every} DH. In particular, not
only does the direct analog of Hawking's area theorem hold on DHs,
but there is a precise \emph{quantitative} relation between the
growth of area of a DH and the amount of energy falling into it
\cite{ak1,ak2}. Therefore, in striking contrast with EHs, we can
associate a direct physical significance to the growth in area of
DHs. This and other quantitative relations have already made DHs
very useful in numerical simulations of black hole formation and
mergers \cite{akrev}. Finally, since they refer only to the
space-time geometry in their immediate vicinity, the existence and
properties of these horizons are insensitive to what happens near
the singularity. Thus, quantum gravity modifications in the
space-time geometry in the vicinity of the classical
singularity would have no effect on these horizons.%
\footnote{One might wonder: Don't the singularity theorems
essentially guarantee that if there is an MTT there must be a
singularity? Recall however that the theorems also assume classical
Einstein's equations and certain energy conditions. Both these
assumptions would be violated in quantum gravity. Therefore, it is
perfectly feasible for MTTs to exist even though the (quantum)
space-time has no singularities.}

Conceptually these quasi-local horizons are also useful in quantum
considerations. Let us first consider equilibrium situations. In
loop quantum gravity, there is a statistical mechanical derivation
of the entropy associated with any isolated horizon
\cite{entropy,alrev}. These cover not only the familiar stationary
black holes but also hairy black holes as well as cosmological
horizons. Next, consider dynamics. During the collapse, the MTT is
space-like and we have a DH. But once the in-fall of matter ends,
the mass of the black hole must decrease and the horizon area must
shrink. In this phase the MTT is time-like and so there is no
obstruction at all for leakage of matter from inside the MTT to the
outside region.

To summarize, black holes were first described using EHs. While this
description has led to important insights, they also have some
important limitations in the dynamical context. The more recent
quasi-local horizons provide concepts and tools that are more
directly useful both in numerical relativity and quantum gravity.

\subsection{Hawking radiation and information loss}
\label{s3.2}

Consider a spherically symmetric gravitational collapse depicted in
Fig \ref{fig:collapse}a. Once the black hole is formed, space-time
develops a new, future boundary at the singularity, whence one can
not reconstruct the geometry and matter fields by evolving the data
\emph{backward} from future null infinity, ${\sp}$. Thus, whereas an
appropriately chosen family of observers near ${\sm}$ has full
information needed to construct the entire space-time, no family of
observers near ${\sp}$ has such complete information. In this sense,
black hole formation leads to information loss. Note that, contrary
to the heuristics often invoked, this phenomenon is not directly
related to black hole uniqueness results: it occurs even when
uniqueness theorems fail, as with `hairy' black holes or in presence
of matter rings non-trivially distorting the horizon. The essential
ingredient is the future singularity which can act as the sink of
information.

A natural question then is: what happens in quantum gravity? Is
there again a similar information loss? Hawking's \cite{swh3}
celebrated work of 1974, mentioned in section \ref{s3.1}, analyzed
this issue in the framework of quantum field theory in curved
space-times. In this approximation, three main assumptions are made:
i) the gravitational field can be treated classically; ii) one can
neglect the back-reaction of the spontaneously created matter on the
space-time geometry; and iii) the matter quantum field under
investigation is distinct from the collapsing matter, so one can
focus just on spontaneous emission. Under these assumptions, at late
times there is a steady emission of particles to ${\sp}$ and the
spectrum is thermal at a temperature dictated by the surface gravity
of the final black hole. In particular, pure states on ${\sm}$
evolve to mixed states on ${\sp}$. However, this external field
approximation is too crude; in particular it violates energy
conservation. To cure this drawback, one can include back-reaction.
A detailed calculation is still not available. However, following
Hawking \cite{swh3}, one argues that, as long as the black hole is
large compared to the Planck scale, the quasi-stationary
approximation should be valid. Then, by appealing to energy
conservation and the known relation between the mass and the horizon
area of \emph{stationary} black holes, one concludes that the area
of the EH should steadily decrease.%
\footnote{This does not contradict the area law because the energy
conditions used in its derivation are violated by the quantum
emission.}
This then leads to black hole evaporation depicted in figure
\ref{fig:collapse}b \cite{swh1}. If one does not examine space-time
geometry but uses instead intuition derived from Minkowskian
physics, one may be surprised that although there is no black hole
at the end, the initial pure state has evolved in to a mixed state.
Note however that even after the inclusion of back reaction, in this
scenario \emph{there is still a final singularity, i.e., a final
boundary in addition to} $\sp$. Therefore, it is not at all
surprising that, in this approximation, information is lost
---it is still swallowed by the final singularity. Thus,
provided figure \ref{fig:collapse}b is a reasonable approximation of
black hole evaporation and one does not add new input `by hand',
then pure states must evolve in to mixed states.

The question then is to what extent this diagram is a good
representation of the physical situation. The general argument in
the relativity community has been the following. Figure
\ref{fig:collapse}b should be an excellent representation of the
actual physical situation as long as the black hole is much larger
than the Planck scale. Therefore, problems, if any, are associated
\emph{only} with the end point of the evaporation process. It is
only here that the semi-classical approximation fails and one needs
full quantum gravity. Whatever these `end effects' are, they deal
only with the Planck scale objects and would be too small to recover
the correlations that have been steadily lost as the large black
hole evaporated down to the Planck scale. Hence pure states must
evolve to mixed states and information is lost.

Tight as this argument seems, it overlooks two important
considerations. First, one would hope that quantum theory is free of
infinities whence figure \ref{fig:collapse}b can not be a good
depiction of physics near the \emph{entire singularity} ---not just
near the end point of the evaporation process. Second, as we saw in
section \ref{s3.1}, the EH is a highly global and teleological
construct. Since the structure of the \emph{quantum} space-time
could be very different from that of figure \ref{fig:collapse}b near
(and `beyond') the singularity, the causal relations implied by the
presence of the EH of figure \ref{fig:collapse}b is likely to be
quite misleading \cite{ph}. Indeed, using the AdS/CFT conjecture,
string theorists have argued that the evolution must be unitary and
information is not lost. However, since the crux of that argument is
based on the boundary theory (which is conjectured to be equivalent
to string theory in the bulk), this line of reasoning does not
provide a direct \emph{space-time description} of how and why the
information is recovered. Where does the above reasoning of
relativists, fail? How must it be corrected?

I believe that answer to these question lies in the fact that,
because of singularity resolution, the quantum space-time is larger
than the classical \cite{ab2}. In support of this view, in the next
two sections I will use a 2-dimensional black hole to argue that the
loss of information is not inevitable even in space-time
descriptions favored by relativists.

\subsection{CGHS black holes}
\label{s3.3}

Let us begin with the spherical collapse of a massless scalar field
$f$ in 4 space-time dimensions resulting in a black hole. Because of
spherical symmetry, it is convenient to factor out by the 2-spheres
of symmetry and pass to the $r-t$ plane. Let us express the
4-dimensional space-time metric $ {}^4\!g_{ab}$ as:
$$ {}^4\!g_{ab}= g_{ab} + \f{e^{-2\phi}}{\kappa^2}\, s_{ab} \, ,
$$
where we have introduced a constant $\kappa$  with dimensions of
inverse length and set $r= e^{-\phi}/\kappa$. Then the (symmetry
reduced) Einstein Hilbert action becomes
\be \label{action1} S(g,\phi,\kappa) = \f{1}{2G}\, \int\! d^2x\,
\sqrt{|g|}\, \big[ e^{-2\phi}\,(R + {\bf 2} \nabla^a\phi\nabla_a
\phi + {\bf 2 e^{-2\phi}} \kappa^2) \,\,+ G\, {\bf
e^{-\phi}}\nabla^a f \nabla_a f\big]  \nonumber \ee
where $R$ is the scalar curvature of the 2-metric $g$.  This theory
is very rich especially because the well-known critical phenomena.
The classical equations cannot be solved exactly. However, an
apparently small modification of this action ---changes in the bold
coefficients (compare (\ref{action1}) and (\ref{action2}))--- gives
a 2-dimensional theory which \emph{is} exactly soluble classically.
There is again a black hole formed by gravitational collapse and it
evaporates by Hawking radiation. This is the Callen, Giddings,
Harvey, Strominger (CGHS) model \cite{cghs} and it arose upon
symmetry reduction of a low energy action motivated by string
theory. Because it has many of the qualitative features of the 4-d
theory but is technically simpler, the model attracted a great deal
of attention in the 90s\,\, (for reviews, see e.g.,
\cite{cghs-rev}). Here, the basic fields are again a 2-dimensional
metric $g$ of signature -+, a geometrical scalar field $\phi$,
called the dilaton, and a massless scalar field $f$. The action is
given by:
\be \label{action2} S(g,\phi,f) := \f{1}{2G}\, \int\! d^2x\,
\sqrt{|g|}\, \big[ e^{-2\phi}\,( R + \mathbf{4} \nabla^a\phi\nabla_a
\phi + \mathbf{4}\kappa^2) + G\, \nabla^a f \nabla_a f\big]
\nonumber \ee
We will analyze this 2-d theory in its own right.
\begin{figure}[]
  \begin{center}
    \includegraphics[width=1.8in,angle=0]{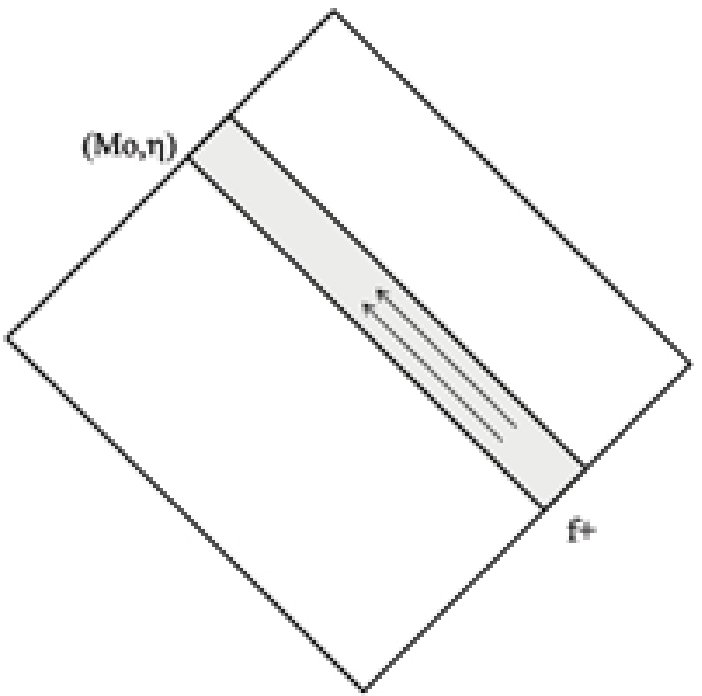} 
    \hskip3cm
\includegraphics[width=1.8in,angle=0]{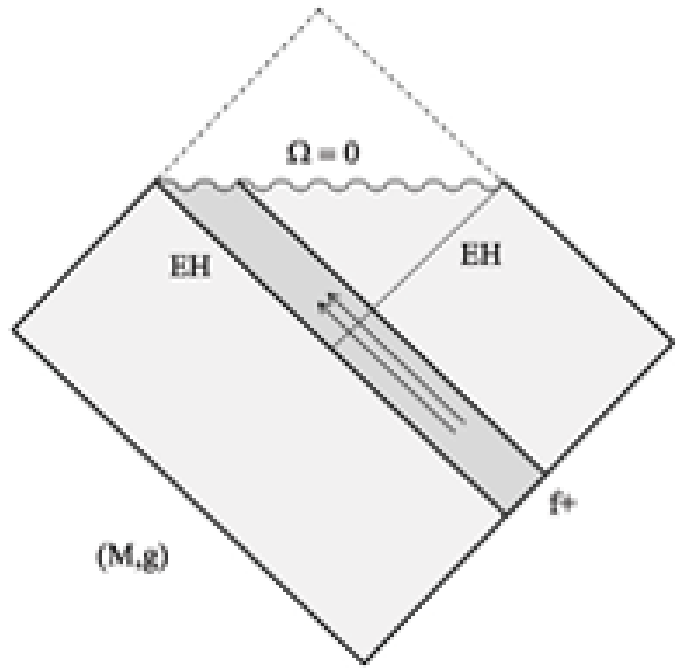}
\caption{$a)$ A typical solution for the $f_+$ mode in Minkowski
space. $b)$ When interpreted in terms of the physical metric $g$,
a black hole has formed because of the gravitational collapse of
$f_+$. The physical space-time $M$ is a proper subset of $M_o$
but the subset realized depends on the solution $f_+$. Therefore
already in the classical Hamiltonian theory, the kinematical
arena is provided by $M_o$.}
\label{cghs-class}
  \end{center}
\end{figure}

Recall that imposition of spherical symmetry in 4-d general
relativity implies that the gravitational field is completely
determined by matter ---the true degrees of freedom are all
contained in the matter. The same is true in the CGHS model.
Furthermore, in the CGHS case there is a simplification: the
equation of motion of $f$ is just $\Box_{(g)} f =0$; \, dynamics of
$f$ is decoupled from $\phi$.\, In 2 dimensions, the physical metric
$g^{ab}$ is conformally related to a flat metric $g^{ab} = \Omega
\eta^{ab}$ and conformal invariance of the wave equation implies
that $\Box_{(g)} f =0$ if and only if  $\Box_{(\eta)} f =0$.
Therefore, we can fix a fiducial flat metric $\eta$, parameterize
$g$ by $\Omega$ and determine $f$ by solving the wave equation on
the 2-dimensional Minkowski space $(M_o,\eta)$. Finally, let us set
$\Phi = e^{-2\phi}$ and write the conformal factor $\Omega$ as
$\Omega = \Theta^{-1}\,  \Phi$. The passage $(g,\phi,f)
\longrightarrow (\Theta, \Phi, f)$ just corresponds to a convenient
choice of field redefinitions.

Since $\Box_{(\eta)}f =0$, we know $f = f_+(z^+) \,+\, f_-(z^-)$,
where $f_\pm$ are arbitrary smooth functions of their arguments and
$z^\pm$ are the advanced and retarded coordinates of $\eta$ (i.e.,
$\eta_{ab} = -\partial_{(a}z^+\, \partial_{b)} z^-$). Given $f$, the
equations of motion for $\Theta$ and $\Phi$ (together with
appropriate boundary conditions) determine the classical solution
completely. To display it, it is simplest to use coordinates $x^\pm$
given by:
$$ \k x^+ = e^{\k z^+} \quad {\rm and} \quad \k x^- = -e^{-\k z^-}\,
.$$
Then, for any given $f_\pm$, the solution is given by
\ba \label{sol1} \Theta &=& - \k^2 x^+\,x^-  \quad {\rm and} \quad
\nonumber\\
\Phi &=& \Theta - \f{G}{2}\textstyle{\int_0^{x^+}} \dd\bar{x}^+\,
\textstyle{\int_0^{\bar{x}^+}} \dd \bar{\bar{x}}^+\, (\partial
f_+/\partial \bar{\bar{x}}^+)^2 \,-\, \f{G}{2}
\textstyle{\int_{0}^{{x}^-}} \dd {\bar{x}}^-\,
\textstyle{\int_{0}^{\bar{x}^-}} \dd\bar{\bar{x}}^-\, (\partial
f_-/\partial \bar{\bar{x}}^-)^2\, .\nonumber\\ \ea
The black hole sector of interest is obtained by setting $f_-=0$ as
in Fig \ref{cghs-class}a and letting $f_+$ collapse. (Alternatively,
one could $f_+=0$ and consider the collapse of $f_-$.)

But why is there a black hole? Fields $f_+, \Theta, \Phi$ are all
smooth on the entire manifold $M_o$. Recall, however, that the
physical metric is given by $g^{ab} = \Omega \eta^{ab} \equiv
\Theta^{-1} \Phi\, \eta^{ab}$. On the entire manifold $M_o$,
$\Theta$ is smooth and nowhere vanishing. However, it is easy to
verify that $\Phi$ vanishes along a space-like line\, (see Fig
\ref{cghs-class}b). On this line $g^{ab}$ becomes degenerate and its
scalar curvature diverges. Thus there is a space-like singularity;
the physical space-time manifold $M$ on which $g_{ab}$ is well
defined is only a part of the fiducial Minkowski manifold $M_o$ (see
Fig \ref{cghs-class}b). Is it hidden behind an event horizon? To ask
this question, we should first verify that $(M, g_{ab})$ admits a
\emph{complete} \cite{gh} future null infinity $\sp$ and the past of
$\sp$ does not contain the singularity. In 2 space-times dimensions,
past as well as future null infinity has two pieces, one to the
right and the other to the left and they are joined only by points
$i^\pm$ at time-like infinity. In the solutions under consideration
$\spr$ is complete but $\spl$ is not. Therefore strictly we can
meaningfully ask if there is a black hole only with respect to
$\spr$ and the answer is in the affirmative. Fortunately to analyze
the Hawking radiation and information loss, we can focus just on
$\spr$. Before going on to these issues, it is interesting to note
that there is a black hole in spite of the fact that the solution
$(f,\theta,\Phi)$ is perfectly regular. This is because the physical
meaning of the solution has to be analyzed using the physical
geometry determined by $g$.

Solution (\ref{sol1}) represents a black hole formed by the
gravitational collapse of $f_+$. In the spirit of Hawking's original
derivation, let us study the dynamics of a \emph{test} quantum field
$\h{f}_-$  on this black hole geometry. Now $\sm$ of every physical
metric $g$ coincides with the past null infinity $\szm$ of Minkowski
space $(M_o, \eta)$ and $g= \eta$ in a neighborhood of $\szml$. So
we can begin with the vacuum state $\ket{0}_-$ at $\szml$ and ask
for its dynamical content. In the Heisenberg picture, the operators
evolve and state remain fixed. The issue then is that of
interpretation of the fixed state $\ket{0}_-$ in the geometry given
by $g$ in a neighborhood of $\spr$. Now, two important factors of
the geometry come into play. First, although the physical metric $g$
is asymptotically flat, \emph{it does not agree with $\eta$ even at}
$\spr$. More precisely, the affine parameter $y^-$ at $\spr$ is a
non-trivial function of $z^-$, reflecting the fact that the
asymptotic time translation of $g$ does not coincide with any of the
asymptotic time translations of $\eta$. Therefore there is a mixing
of positive and negative frequency modes. Since $\ket{0}_-$ is
defined using $z^-$, it is populated with particles defined at
$\spr$ by $g$. Second, $\spr$ is a proper subset of $\szpr$.
Therefore, we have to trace over modes of $\h{f}_-$ with support on
$\szpr\, -\, \spr$. Therefore, as far as measurements of observables
near $\spr$ are concerned, the state $\ket{0}_-$ is
indistinguishable from a density matrix $\rho$ on the Hilbert space
$\H$ of $\h{f}_-$ at $\spr$. Detailed calculation shows that at late
times, $\rho$ is precisely the thermal state at temperature
$\hbar\kappa/2\pi$ \cite{gn}! Thus, in the CGHS model there is
indeed Hawking radiation and therefore, by repeating the reasoning
summarized in section \ref{s3.2} one can conclude that there must be
information loss.

I will conclude this section by summarizing the similarities and
differences in the 4 and 2 dimensional analyses. In both cases there
is a formation of a black hole due to gravitational collapse and the
test quantum field is distinct from the field that collapses. Thanks
to asymptotic flatness at past null infinity, the vacuum state
$\ket{0}_-$ of the test field is well-defined and the key issue is
that of its physical interpretation in the physical geometry near
future null infinity. Finally although $\kappa$ was introduced as a
constant in the CGHS theory, one can verify that it is in fact the
surface gravity of the stationary black hole in the future of the
support of $f_+$. In both cases the Hawking temperature is this
given by $\hbar/2\pi$ times the surface gravity. However, there are
also some important differences. First, whereas there is just one
$\sm$ and $\sp$ in 4 dimensions in the CGHS case we have two copies
of each and the clear-cut black hole interpretation holds only with
respect to $\spr$. Second, while in 4 dimensions $\kappa$ and hence
the Hawking temperature is inversely proportional to the mass of the
black hole, in the CGHS case it is a constant. Finally, at a
technical level, even in the spherically symmetric reduction of the
4-dimensional theory, the equation satisfied by the scalar field $f$
is much more complicated than the CGHS wave equation. Therefore,
while analysis of the CGHS black hole does provide valuable insights
for the 4 dimensional case, one cannot take directly over results.

\subsection{Quantum geometry}
\label{s3.4}

Since the model is integrable classically, many steps in the passage
to quantum theory are simplified \cite{atv}. Our basic fields will
again be $\h{f}, \h\Theta, \h\Phi$. The true degree of freedom is in
the scalar field $f$ and it satisfies just the wave equation on
Minkowski space $(M_o,\eta)$. Therefore, it is straightforward to
construct the Fock space $\F = \F_+\otimes \F_-$ and represent
$\h{f}_\pm$ as operator valued distributions on $\F$. Classically,
we have explicit expressions (\ref{sol1}) of fields $\Theta$ and
$\Phi$ in terms of $f$ on all of $M_o$. In quantum theory, because
of trace anomaly the equations satisfied by $\h\Theta,\h{\Phi}$ are
more complicated. Therefore explicit solutions are not available.
However, these are hyperbolic equations on the fiducial Minkowski
space and the boundary values \emph{at} $\szm$ are given by the
(unambiguous) operator versions of (\ref{sol1}). Therefore, in
principle, it should be possible to solve them. A conjecture based
on approximate solutions is that $\h\Theta$ would be an operator
field and $\h\Phi$ an operator valued distribution on $\F$.

At first, may seem surprising that there is no Hilbert space
corresponding to geometry. However, already at the classical level
the covariant phase space can be coordinatized completely by the
scalar field $f$ and geometric fields $\Theta, \Phi$ are just
functionals on this phase space. The situation in quantum theory is
precisely what one would expect upon quantization. While the full
quantum theory is still incomplete in the CGHS model, there is a
simpler and interesting system in which this feature is realized in
detail: cylindrical gravitational waves in 4-dimensional general
relativity. This system is equivalent to 2+1 Einstein gravity
coupled to an axi-symmetric scalar field. Again because there are no
gravitational degrees of freedom in 2+1 dimensions, the true degree
of freedom can be encoded in the scalar field which now satisfies a
wave equation in a fiducial 2+1 dimensional Minkowski space. The
regulated metric operator is represented as an operator valued
distribution on the Fock space of the scalar field \cite{ap} and
leads to interesting and unforeseen quantum effects \cite{aa-large}.

Returning to the CGHS model, we can now ask: What is a quantum black
hole? In the classical theory, black holes result if we specify a
smooth profile $f^o$ as initial data for $f_+$ on $\szmr$ and zero
data for $f_-$ on $\szml$. In quantum theory, then, a candidate
black hole would a quantum state $\ket{\Psi}$ which is peaked at
this classical data on $\szm$: $\ket{\Psi} = \ket{0}_- \otimes
\ket{C_f^{o}}_+$ where $\ket{0}_-$ is the vacuum state in $\F_-$ and
$\ket{C_f^{o}}_+$ is the coherent state in $\F_+$ peaked at the
classical profile $f_o$ of $f_+$.  One can show that if one solves
the quantum equations for $\h{\Theta},\h{\Phi}$ in a certain
approximation (the 1st step in a certain bootstrapping), then the
states $\Psi$ do emerge as black holes: the expectation values of
$\hat{g}^{ab}, \h{\Phi}$ are precisely those of the classical black
hole solutions. In particular, $\langle \h\Phi \rangle$ vanishes
along a space-like line which appears as the singularity in the
classical theory. However, the true \emph{quantum} geometry near
this classical singularity is perfectly regular \cite{atv}: the
\emph{operator} $\h\Phi$ does not vanish, only its expectation value
does. Furthermore, one can also calculate fluctuations and show that
they are small near infinity but huge near the classical
singularity. Consequently, the expectation values are poor
representations of the actual quantum geometry in a neighborhood of
the classical singularity. The fact that the quantum metric
$\hat{g}^{ab}$ is regular on $M_o$ already in this approximation
suggests that the singularity may be resolved in the quantum theory
making the quantum space-time larger than the classical one. There
is then a possibility that there may be no information loss.

\begin{figure}[]
  \begin{center}
    \includegraphics[width=1.8in,angle=0]{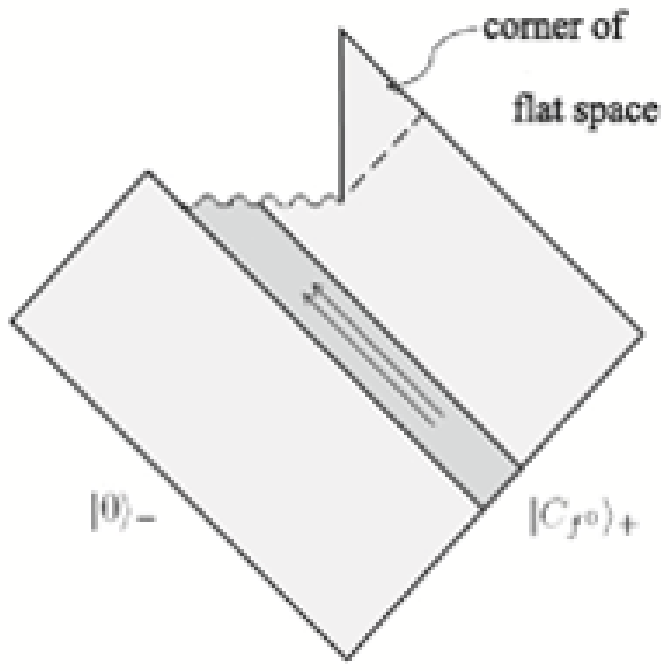}
    \hskip3cm
\includegraphics[width=1.8in,angle=0]{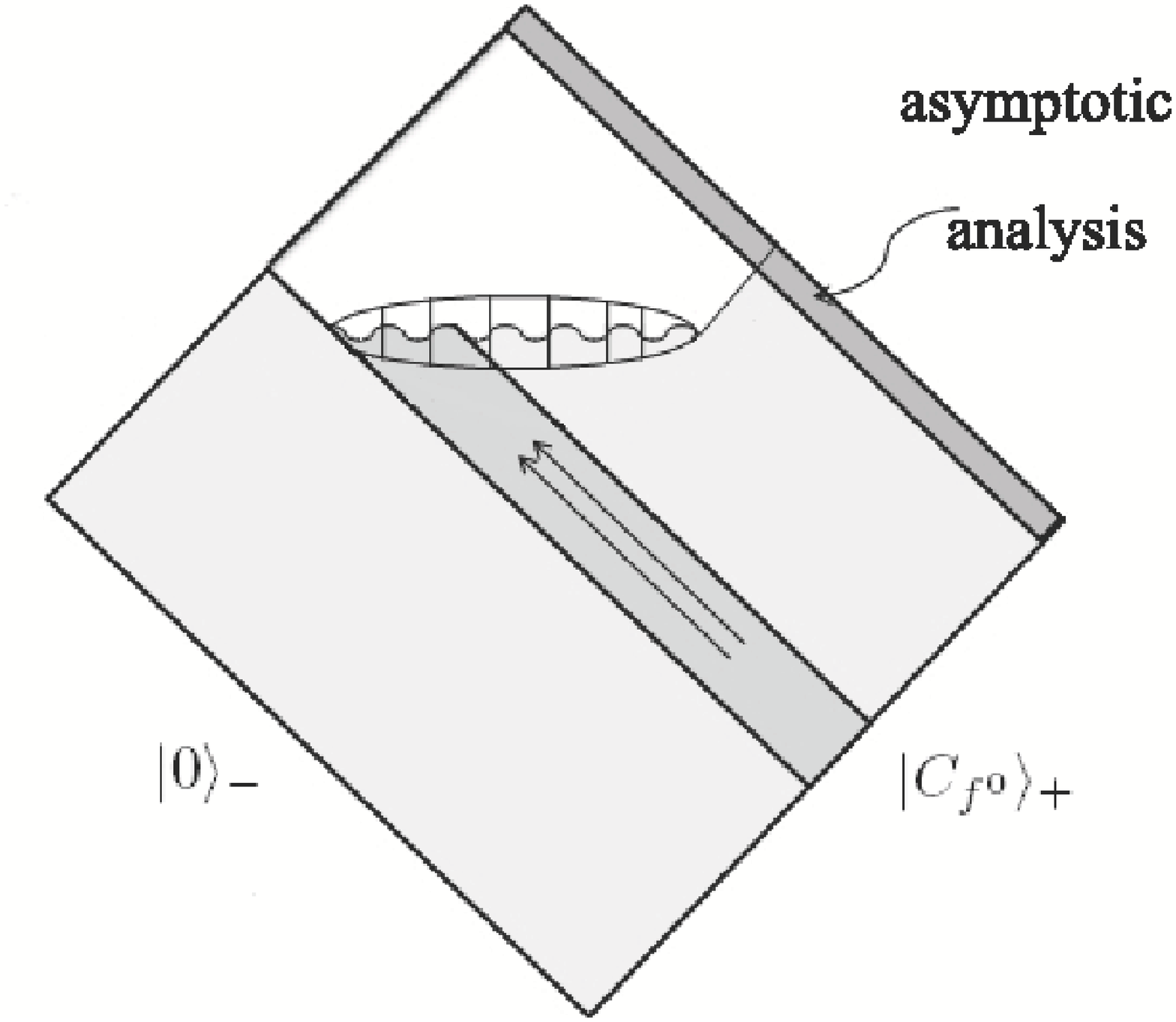}
\caption{$a)$ The CGHS analog of the Penrose diagram
\ref{fig:collapse}b. This diagram has been used for a number
of years to describe space-time geometry after inclusion
of back reaction. Singularity is still part of the future
boundary and so the information is lost. $b)$ The space-time
diagram suggested by the asymptotic analysis of mean field
equations near $\spr$. In the quantum space-time $\spr$ is
`as long as' $\szpr$, whence $\ket{0}_-$ is a pure state also
with respect to the physical metric $g$. It is however populated
by particles and resembles thermal density matrix at an
intermediate region of $\spr$.}
 \label{cghs-quantum}
  \end{center}
\end{figure}

This issue is probed using the mean field approximation (MFA)
\cite{atv}. Here, one first takes the expectation value of the the
quantum equations governing  $\h{\Theta},\h{\phi}$ in the state
$\Psi$ and, furthermore, replaces $\Phi$, $\theta$ by their
expectation values. Thus, for example,
$\langle\h\Theta\,\h\Phi\rangle$ is replaced by
$\langle\h\Theta\rangle\,\, \langle \h\Phi\rangle$ but $\langle
:(\partial \h{f})^2: \rangle$ is kept as is. This amounts to
ignoring the quantum fluctuations in the geometric operators
$\h\Theta, \h\Phi$ but not those in the matter field $\h{f}$.  This
approximation can be justified in the limit in which there is a
large number $N$ of scalar fields $\h{f}$ rather than just one and
we restrict ourselves to regions in which ``fluctuations in the
geometry are less than $N$ times the fluctuations in any one matter
field''. In this region, the mean field approximation provides a
good representation of the geometry that includes back reaction of
the Hawking radiation.

It turns out that the resulting equations on $\bar\Theta := \langle
\h\Theta\rangle$ and $\bar\Phi := \langle \h\Phi\rangle$ were
already obtained sometime ago using functional integral techniques
and solved numerically \cite{num}. By making appeal to the
4-dimensional theory whose symmetry reduction gives the CGHS models,
one can introduce the notion of marginally trapped `surfaces' and
their `area'. Simulations showed that marginally trapped surfaces do
form due to infalling matter, the marginally trapped tube is first
space-like ---i.e., is a dynamical horizon--- but, after the inflow
of collapsing matter ends, becomes time-like due to the leakage of
the Hawking radiation. Thus the scenario based on quasi-local
horizons is realized. In the dynamical horizon phase, the horizon
area $a_\hor$ increases and in the subsequent Hawking evaporation,
it decreases to zero: It is again the MTT that evaporates. However,
further evolution to the future moves one closer to what was the
classical singularity. As I mentioned above, in this region the
quantum fluctuations in geometry become huge and so the mean field
approximation fails. The simulations cannot be continued further.
However, since the area of the marginally trapped surface shrunk to
zero, it was assumed ---as is reasonable--- that the Bondi mass at
the corresponding retarded instant of time would be zero on $\spr$.
Therefore, following what Hawking did in 4-dimensions, it became
customary to attach by hand a `corner of Minkowski space' to the
numerically evolved space-time thereby arriving at a Penrose diagram
of figure \ref{cghs-quantum}a. Note that in this diagram, the future
boundary for the $\h{f}_-$ modes consists not just of $\spr$ but
also a piece of the singularity. As I argued in section \ref{s3.2},
if this is an accurate depiction of the physical situation, one
would conclude that $\ket{0}_-$ at $\szml$ would evolve to a density
matrix on $\spr$ and information would indeed be lost.

Note however that the key to the information loss issue lies in the
geometry near future infinity and MFA should be valid there. Thus,
rather than attaching a corner of flat space by hand at the end of
the numerical simulation, we can use the mean field equations near
$\spr$ and let them tell us what the structure of $\spr$ of the
\emph{physical} metric is.

To realize this idea, one has to make three assumptions: i) exact
quantum equations can be solved and the expectation value
$\bar{g}^{ab}$ of $\hat{g}^{ab}$ admits a smooth right null infinity
$\sbpr$ which coincides with $\szpr$ \emph{in the distant past}
(i.e. near $i^o_{\rm R}$);\, ii) MFA holds in a neighborhood of
$\sbpr$; and, \, iii) Flux of quantum radiation vanishes at some
finite value of the affine parameter ${y}^-$ of $\sbpr$ defined by
the asymptotic time translation of $\bar{g}$. All three assumptions
have been made routinely in the analysis of the information loss
issue, although they are often only implicit. Indeed, one cannot
even meaningfully ask if information is lost unless the first two
hold. (The third assumption can be weakened to allow the flux to
decay sufficiently fast in the future.) Then, a systematic analysis
of the MFA equations shows \cite{atv} that \emph{the right future
null infinity $\spr$ of the physical metric $\bar{g}$ coincides with
that of} $\eta$; \,\, $\sbpr = \szpr$\, (see Fig
\ref{cghs-quantum}). This implies that to interpret $\ket{0}_-$ at
$\sbpr$ we no longer have to trace over any modes; in contrast to
the situation encountered in the external field approximation
discussed in section \ref{s3.3}, all modes of $\h{f}_-$ are now
accessible to the asymptotically stationary observers of $\bar{g}$.
The vacuum state $\ket{0}_-$ of $\eta$ is pure also with respect to
$\bar{g}$. But is it in the asymptotic Fock space of $\bar{g}$?
Calculation of Bogoluibov coefficients shows \cite{atv} that the
answer is in the affirmative. Thus, the interpretation of
$\ket{0}_-$ with respect to $\bar{g}$ is that it is a pure state
populated by pairs of particles at $\sbpr$. \emph{There is neither
information loss nor remnants.}

Let us summarize the discussion of CGHS black holes. A key
simplification in this model is that the matter field satisfies just
the wave equation on $(M_o,\eta^{ab})$. Therefore, given initial
data on $\szm$, we already know the state everywhere both in the
classical and the quantum theory. However, the state derives its
physical interpretation from geometry which is a complicated
functional of the matter field. We do not yet know the quantum
geometry everywhere. But approximation methods suggest that
$\hat{g}^{ab}$ is likely to be well-defined (and nowhere vanishing)
everywhere on $M_o$.  By making rather weak assumptions on the
asymptotic behavior of its expectation value $\bar{g}^{ab}$, one can
conclude that the right future null infinity $\sbpr$ of
$\bar{g}^{ab}$ coincides with $\szpr$ of $\eta^{ab}$ and the affine
parameters $y^-$ and $z^-$ defined by the two metrics are such that
the exact quantum state $\ket{0}_-$ is a pure state in the
asymptotic Fock space of $\bar{g}^{ab}$. The S-matrix is unitary and
there is no information loss. Thus the asymptotic analysis leads us
to a Penrose diagram of Fig \ref{cghs-quantum}b which is
significantly different from Fig \ref{cghs-quantum}a,  based on
Hawking's original proposal \cite{swh3}. In particular, the quantum
space-time does not end at a future singularity and is larger than
that in \ref{cghs-quantum}a. The singularity is replaced by a
genuinely quantum region in which quantum fluctuations are large and
the notion of a smooth metric tensor field is completely inadequate.
However, in contrast to the situation in quantum cosmology of
section \ref{s2}, a full solution to the quantum equations is still
lacking.

\section{Discussion}
\label{s4}

In section \ref{s2} we saw that many of the long standing questions
regarding the big bang have been answered in detail in the FRW
cosmologies with a massless scalar field and the results are
physically appealing. Main departures from the \WDW theory occur due
to \emph{quantum geometry effects} of LQG. There is no fine tuning
of initial conditions, nor a boundary condition at the singularity,
postulated from outside. Also, there is no violation of energy
conditions. Indeed, quantum corrections to the matter Hamiltonian do
not play any role in the resolution of singularities of these
models. The standard singularity theorems are evaded because the
geometrical side of the classical Einstein's equations is modified
by the quantum geometry corrections of LQC. While the detailed
results presented in section \ref{s2.5} are valid only for these
simplest models, partial results have been obtained also in more
complicated models indicating that the singularity resolution is
rather robust.

In this respect there is a curious similarity with the very
discovery of physical singularities in general relativity. They were
first encountered in special examples. But the examples were also
the physically most interesting ones ---e.g., the big-bang and the
Schwarzschild curvature singularities. At first it was thought that
these space-times are singular because they are highly symmetric. It
was widely believed that generic solutions of Einstein'e equations
should be non-singular. As is well-known, this belief was shattered
by the Penrose-Hawking singularity theorems. Some 40 years later we
have come to see that the big bang and the Schwarzschild
singularities are in fact resolved by quantum geometry effects. Is
this an artifact of high symmetry? Or, are there robust
\emph{singularity resolution theorems} lurking just around the
corner?

A qualitative picture that emerges is that the non-perturbative
quantum geometry corrections are \emph{`repulsive'}.%
\footnote{We saw in section \ref{s2.4} that there is no connection
operator in LQG. As a result the curvature operator has to be
expressed in terms of holonomies and becomes non-local. The
repulsive force can be traced back to this non-locality.
Heuristically, the polymer excitations of geometry do not like to be
packed too densely; if brought too close, they repel.}
While they are negligible under normal conditions, they dominate
when curvature approaches the Planck scale and can halt the collapse
that would classically have lead to a singularity. In this respect,
there is a curious similarity with the situation in the stellar
collapse where a new repulsive force comes into play when the core
approaches a critical density, halting further collapse and leading
to stable white dwarfs and neutron stars. This force, with its
origin in the Fermi-Dirac statistics, is \emph{associated with the
quantum nature of matter}. However, if the total mass of the star is
larger than, say, $5$ solar masses, classical gravity overwhelms
this force. The suggestion from LQC is that a new repulsive force
\emph{associated with the quantum nature of geometry} comes into
play and is strong enough to counter the classical, gravitational
attraction, irrespective of how large the mass is. It is this force
that prevents the formation of singularities. Since it is negligible
until one enters the Planck regime, predictions of classical
relativity on the formation of trapped surfaces, dynamical and
isolated horizons would still hold. But assumptions of the standard
singularity theorems would be violated. There would be no
singularities, no abrupt end to space-time where physics stops.
Non-perturbative, background independent quantum physics would
continue.

One can also analyze the CGHS models using LQG \cite{alok}. However,
I used the more familiar Fock spaces to illustrate the fact that the
basic phenomenon of singularity resolution by quantum geometry
effects is more general. In the CGHS case the analysis is not as
complete as in the cosmological models because the CGHS model has an
infinite number of degrees of freedom. But results obtained using
various approximations strongly suggest that, as in the cosmological
case, quantum space-times are larger than what the classical theory
suggests. However, nature of the quantum space-time is quite
different in the two cases. In the cosmological case the state
remains sharply peaked around a smooth geometry even near the
bounce. The expression of the effective metric which provides an
excellent approximation to the exact quantum state does have an
explicit dependence on $\hbar$ due to quantum corrections. However,
it is smooth everywhere. In the CGHS model on the other hand quantum
fluctuations in the metric become large in the Planck regime whence
one cannot approximate the quantum geometry by \emph{any} smooth
geometry. Rather, there is a genuine quantum bridge joining the
smooth metric in the distant past to that in the distant future.

At first one might think that, since quantum gravity effects concern
only a tiny region, whatever quantum effects there may be, their
influence on the global properties of space-time should be
negligible whence they would have almost no bearing on the issue of
the Beginning and the End. However, as we saw, once the singularity
is resolved, vast new regions appear on the `other side' ushering in
new possibilities that were totally unforeseen in the realm of
Minkowski and Einstein. Which of them are realized generically? Is
there a manageable classification? If, as in the CGHS case, there
are domains in which geometry is truly quantum, classical causality
would be rendered inadequate to understand the global structure of
space-time. Is there a well-defined but genuinely quantum notion of
causality which reduces to the familiar one on quantum states which
are sharply peaked on a classical geometry? Or, do we just abandon
the idea that space-time geometry dictates causality and formulate
physics primarily in relational terms? There is a plethora of such
exciting challenges. Their scope is vast,  they force us to
introduce novel concepts and they lead us to unforeseen territories.
These are just the type of omens that foretell the arrival of a
major paradigm shift to take us beyond the space-time continuum of
Minkowski and Einstein.

\bigskip

\textbf{Acknowledgments:} Much of this chapter is based on joint
work with Alex Corichi, Tomasz Pawlowski, Param Singh, Victor
Taveras, Madhavan Varadarajan and Kevin Vandersloot. Innumerable
discussions with them sharpened my understanding. I have also
benefited greatly from comments, suggestions and probing questions
of many colleagues, especially Martin Bojowald, James Hartle, Gary
Horowitz, Jerzy Lewandowski, Donald Marolf, Roger Penrose and Carlo
Rovelli. This work was supported in part by the NSF grants
PHY04-56913 and PHY05-54771, the Alexander von Humboldt Foundation,
the The George A. and Margaret M. Downsbrough Endowment and the
Eberly research funds of Penn State.


%

\end{document}